\documentclass{article}
\usepackage{times,graphicx}
\begin{document}

\title{A Numerical Testbed for Hypotheses of Extraterrestrial Life and Intelligence}
\author{Duncan H. Forgan}
\maketitle

\noindent Scottish Universities Physics Alliance (SUPA) \\
Institute for Astronomy, University of Edinburgh \\
Royal Observatory Edinburgh \\
Blackford Hill \\
Edinburgh EH9 3HJ \\
UK \\\\
Tel: 0131 668 8359 \\
Fax: 0131 668 8416 \\
Email: dhf@roe.ac.uk \\

\begin{abstract}

\noindent The search for extraterrestrial intelligence (SETI) has been heavily influenced by solutions to the Drake Equation, which returns an integer value for the number of communicating civilisations resident in the Milky Way, and by the Fermi Paradox, glibly stated as: ``If they are there, where are they?''.  Both rely on using average values of key parameters, such as the mean signal lifetime of a communicating civilisation.  A more accurate answer must take into account the distribution of stellar, planetary and biological attributes in the galaxy, as well as the stochastic nature of evolution itself.  This paper outlines a method of Monte Carlo realisation which does this, and hence allows an estimation of the distribution of key parameters in SETI, as well as allowing a quantification of their errors (and the level of ignorance therein).  Furthermore, it provides a means for competing theories of life and intelligence to be compared quantitatively. \\

\noindent Keywords: Numerical, Monte Carlo, extraterrestrial intelligence, SETI, Drake Equation, Fermi Paradox

\end{abstract} 

\newpage

\section{Introduction}

\noindent The science of SETI has always suffered from a lack of quantitative substance (purely resulting from its reliance on one-sample statistics) relative to its sister astronomical sciences. In 1961, Frank Drake took the first steps to quantifying the field by developing the now-famous Drake Equation, a simple algebraic expression which provides an estimate for the number of communicating civilisations in the Milky Way.  Unfortunately, its simplistic nature leaves it open to frequent re-expression, hence there are in fact many variants of the equation, and no clear canonical form.  For the purpose of this paper, the following form will be used (Walters, Hoover \& Kotra (1980) \cite{Drake_eqn}):

\begin{equation} N = R_*f_gf_pn_ef_lf_if_cL \end{equation}

\noindent With the symbols having the following meanings: \\

\noindent \(N\) = The number of Galactic civilisations who can communicate with Earth \\
\noindent \(R_*\) = The mean star formation rate of the Milky Way \\
\noindent \(f_g\) = The fraction of stars that could support habitable planets \\
\noindent \(f_p\) = The fraction of stars that host planetary systems \\
\noindent \(n_e\) = The number of planets in each system which are potentially habitable \\
\noindent \(f_l\) = The fraction of habitable planets where life originates and becomes complex \\
\noindent \(f_i\) = The fraction of life-bearing planets which bear intelligence \\
\noindent \(f_c\) = The fraction of intelligence bearing planets where technology can develop \\
\noindent \(L\) = The mean lifetime of a technological civilisation within the detection window \\

\noindent The equation itself does suffer from some key weaknesses: it relies strongly on mean estimations of variables such as the star formation rate; it is unable to incorporate the effects of the physico-chemical history of the galaxy, or the time-dependence of its terms.  Indeed, it is criticised for its polarising effect on ``contact optimists'' and ``contact pessimists'', who ascribe very different values to the parameters, and return values of \(N\) between \(10^{-5}\) and \(10^6\) (!). \\ 

\noindent A decade before, Enrico Fermi attempted to analyse the problem from a different angle, used order of magnitude estimates for the timescales required for an Earthlike civilisation to arise and colonise the galaxy to arrive at the conclusion that the Milky Way should be teeming with intelligence, and that they should be seen all over the sky.  This lead him to pose the Fermi Paradox, by asking, ``Where are they?''.  The power of this question, along with the enormous chain of events required for intelligent observers to exist on Earth to pose it, has lead many to the conclusion that the conditions for life to flourish are rare, possibly even unique to Earth (Ward and Brownlee (2000)\cite{rare_Earth}.  The inference by Lineweaver (2001)\cite{Line_planets} that the median age of terrestrial planets in the Milky Way is \(1.8 \pm 0.9\) Gyr older than Earth would suggest that a significant number of Earthlike civilisations have had enough time to evolve, and hence be detectable: the absence of such detection lends weight to the so-called ``rare Earth'' hypothesis.  However, there have been many posited solutions to the Fermi Paradox that allow ETI to be prevalent, such as:

\begin{itemize}
\item They are already here, in hiding
\item Contact with Earth is forbidden for ethical reasons
\item They were here, but they are now extinct
\item They will be here, if Mankind can survive long enough
\end{itemize}

\noindent Some of these answers are inherently sociological, and are difficult to model.  Others are dependent on the evolution of the galaxy and its stars, and are much more straightforward to verify. As a whole, astrobiologists are at a tremendous advantage in comparison with Drake and Fermi: the development of astronomy over the last fifty years - in particular the discovery of the first extra solar planet (Mayor et al 1995 \cite{Mayor}) and some hundreds thereafter, as well as the concepts of habitable zones, both stellar (Hart 1979\cite{Hart_HZ}) and galactic (Lineweaver et al 2004 \cite{Line_GHZ}) - have allowed a more in-depth analysis of the problem.  However, the key issue still affecting SETI (and astrobiology as a whole) is that there is no consensus as to how to assign values to the key \emph{biological} parameters involved in the Drake Equation and Fermi Paradox.  Furthermore, there are no means of assigning confidence limits or errors to these parameters, and therefore no way of comparing hypotheses for life (e.g. panspermia - review, see Dose (1986)\cite{panspermia} -  or the ``rare Earth'' hypothesis (Ward and Brownlee (2000)\cite{rare_Earth}).   This paper outlines a means for applying Monte Carlo Realisation techniques to investigate the parameter space of intelligent civilisations more rigorously, and to help assign errors to the resulting distributions of life and intelligence. \\

\noindent The paper is organised as follows: in section \ref{sec:Method} the techniques are described; in section \ref{sec:Inputs} the input data is discussed; in section \ref{sec:Results} the results from several tests are shown, and in section \ref{sec:Conclusions} the method is reviewed.

\section{Method}\label{sec:Method}

\noindent The overall procedure can be neatly summarised as:

\begin{enumerate}
\item Generate a galaxy of \(N_*\) stars, with parameters that share the same distribution as observations
\item Generate planetary systems for these stars
\item Assign life to some of the planets depending on their parameters (e.g. distance from the habitable zone)
\item For each life-bearing planet, follow life's evolution into intelligence using stochastic equations
\end{enumerate}

\noindent This will produce one Monte Carlo Realisation (MCR) of the Milky Way in its entirety.  The concept of using MCR techniques in astrobiology is itself not new: recent work by Vukotic \& Cirkovic (2007,2008)\cite{Cirkovic_time, Cirkovic_neo} uses similar procedures to investigate timescale forcing by global regulation mechanisms such as those suggested by Annis (1999)\cite{Annis}.  In order to provide error estimates, this procedure must be repeated many times, so as to produce many MCRs, and to hence produce results with a well-defined sample mean and sample standard deviation.  The procedure relies on generating parameters in three categories: stellar, planetary and biological.

\subsection{Stellar Properties}

\begin{figure}
\begin{center}
\includegraphics[scale = 0.5]{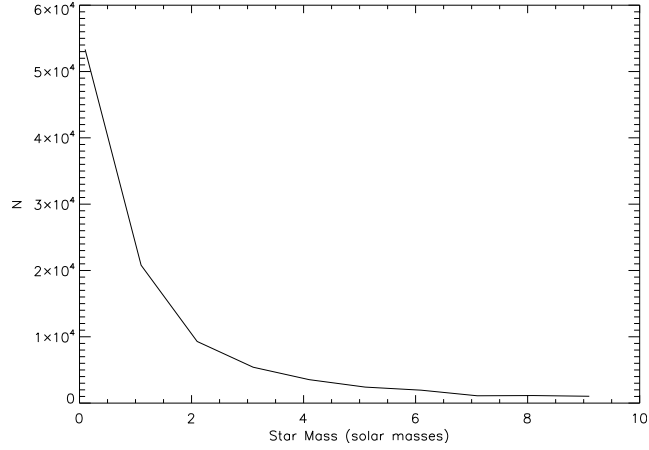}
\caption{\emph{The stellar IMF used in this work (Scalo \& Miller (1979) \cite{IMF}).  This is an example produced as part of a single MCR.}\label{fig:IMF}}
\end{center}
\end{figure}

\noindent The study of stars in the Milky Way has been extensive, and their properties are well-constrained.  Assuming the stars concerned are all main sequence objects allows much of their characteristics to be determined by their mass.  Stellar masses are randomly sampled to reproduce the observed initial mass function (IMF) of the Milky Way, which is taken from Scalo \& Miller (1979)\cite{IMF} (see \textbf{Figure \ref{fig:IMF}}). \\

\noindent  Stellar radii can then be calculated using (Prialnik 2000 \cite{Prialnik}): 

\begin{equation} \frac{R_*}{R_{\odot}} = \left[\frac{M_*}{M_{\odot}}\right]^{\frac{n-1}{n+3}} \end{equation}

\noindent where \(n=4\) if the primary fusion mechanism is the p-p chain (\(M_* \leq 1.1\,M_{\odot}\)), and \(n=16\) if the primary fusion mechanism is the CNO cycle (\(M_* > 1.\,1M_{\odot}\)).  (Please note that in this paper, the subscript \(\odot\) denotes the Sun, e.g. \(M_{\odot}\) indicates the value of one solar mass).  The luminosity is calculated using a simple mass-luminosity relation:

\begin{equation}\frac{L_*}{L_{\odot}} = \left[\frac{M_*}{M_{\odot}}\right]^{3} \end{equation}

\noindent The main sequence lifetime therefore is:

\begin{equation}\frac{t_{MS}}{t_{MS,\odot}} =\left[\frac{M_*}{M_{\odot}}\right]^{-2}  \end{equation}

\noindent The stars' effective temperature can be calculated, assuming a blackbody:

\begin{equation}  T_* = \left[\frac{L_*}{4\pi R_*^2\sigma_{SB}}\right]^{1/4}   \end{equation}

\begin{figure}
\begin{center}
\includegraphics[scale = 0.5]{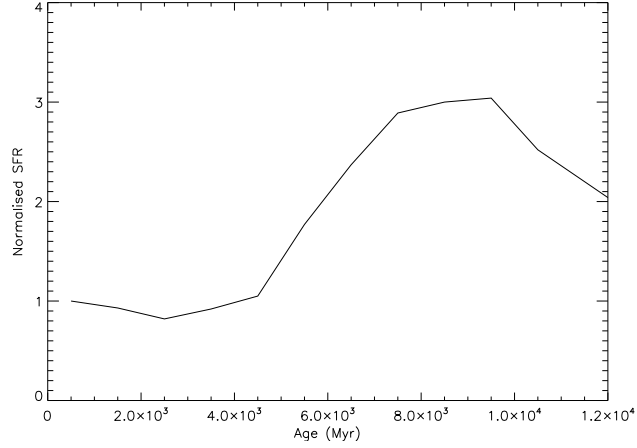}
\caption{\emph{The star formation history used in this work (Twarog (1980) \cite{SFH}} \label{fig:SFH}}
\end{center}
\end{figure}

\noindent The star's age is sampled to reproduce the star formation history of the Milky Way (Twarog 1980 \cite{SFH}), see \textbf{Figure \ref{fig:SFH}}. \\

\noindent The metallicity of the star is dependent on the metallicity gradient in the galaxy, and hence its galactocentric radius, \(r_{gal}\).  This is sampled so that the surface mass density of the galaxy is equivalent to that of the Milky Way (assuming a simple two-dimensional disc structure):

\begin{equation} \Sigma(r_{gal}) = \Sigma_0 e^{-r_{gal}/r_h }  \end{equation} 
 
\noindent where \(r_h\) is the galactic scale length (taken to be 3.5 kpc).   Therefore, given its galactocentric radius, the metallicity of the star, in terms of \(\left[\frac{Fe}{H}\right]\),  is calculated using the simple parametrisation of the abundance gradient:

\begin{equation} Z_* = -z_{grad}\log\left(\frac{r_{gal}}{r_{gal,\odot}}\right) \end{equation}

\begin{figure}
\begin{center}
\includegraphics[scale = 0.5]{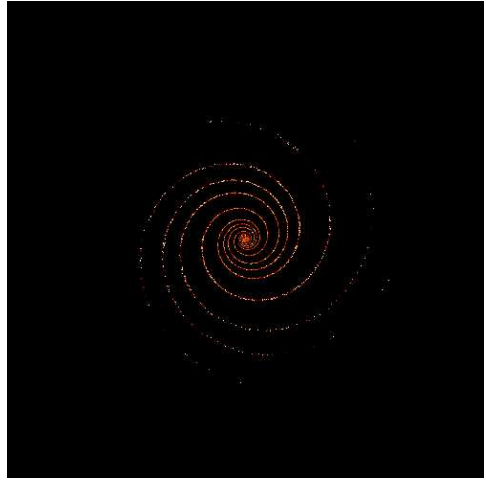}
\caption{\emph{An example star map, taken from one MCR.  Brighter areas indicate greater stellar mass}\label{fig:starmap}}
\end{center}
\end{figure}

\noindent Where \(z_{grad}\) is taken to be 0.07 (Hou, Prantzos and Boissier (2000) \cite{z_grad}). The star is then placed at its galactocentric radius into one of four logarithmic spiral arms (mimicking the four main logarithmic spiral arms of the Milky Way, see \textbf{Figure \ref{fig:starmap}}). \\ 

\begin{figure}
\begin{center}
\includegraphics[scale = 0.5]{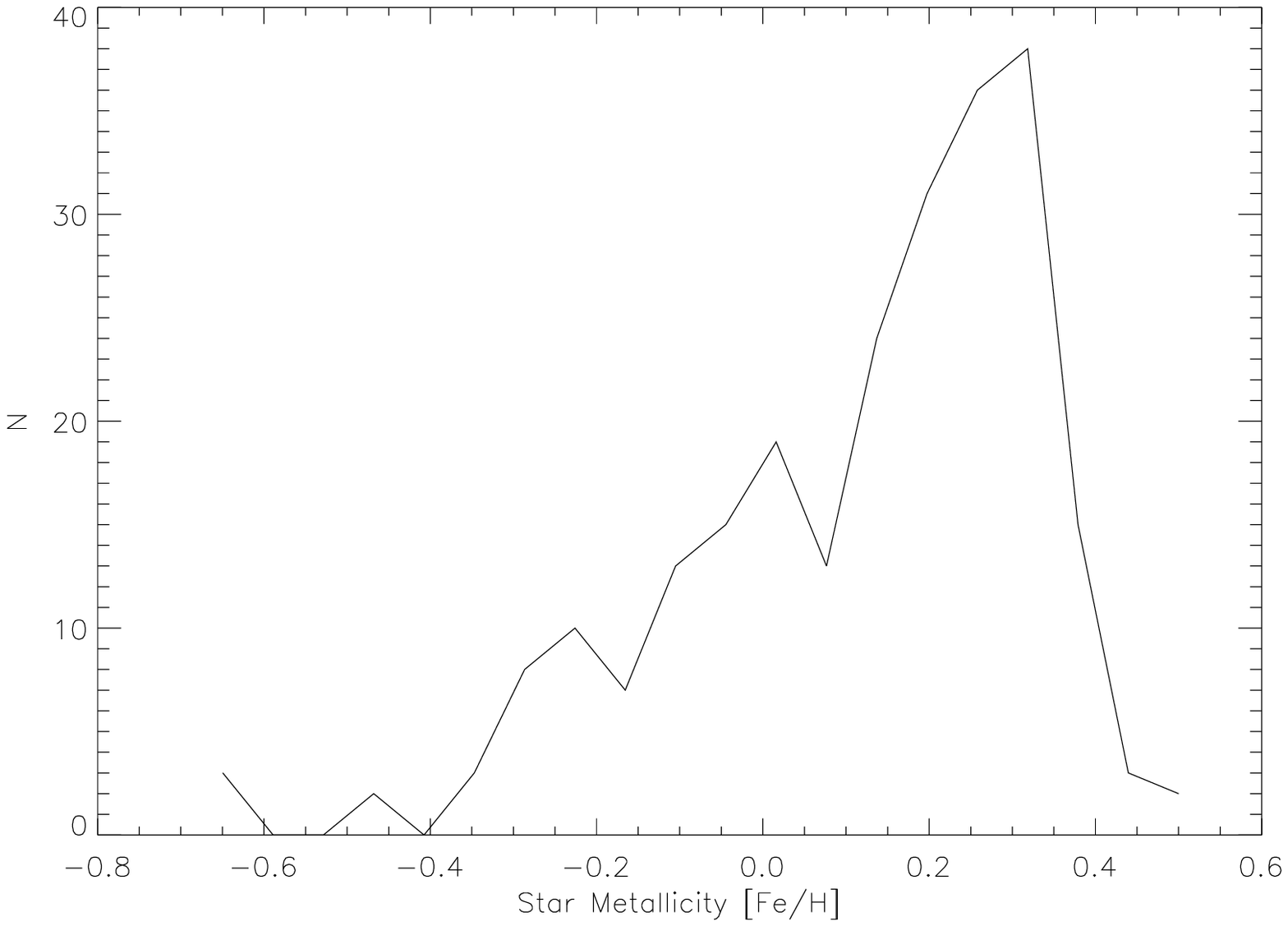}
\caption{\emph{The distribution of host star metallicity, constructed from the Exoplanet Encyclopedia data (http://exoplanet.eu)} \label{fig:p_planet}}
\end{center}
\end{figure}

\noindent  Finally, stars are assigned planets based on their metallicity (sampling the current metallicity relations for extrasolar planets, see \textbf{Figure \ref{fig:p_planet}}).  Some stars will have no planets, some will have only one, and some may have multiple systems (based on the multiplicity of current extrasolar planetary systems)\footnote{Although not considered here, it should be noted that too high a metallicity may be as negative a factor as low metallicity, as high metallicity systems will tend to produce ocean worlds (Leger et al 2004 \cite{ocean_worlds}), which may prove hostile to the formation of intelligent life.}.

\subsection{Planetary Properties}

\noindent Since the discovery of 51 Peg B (Mayor et al 1995 \cite{Mayor}), the data garnered on exoplanets has grown at an increasing rate.  At the time of writing, there are over 300 known exoplanets, discovered with a variety of observational techniques (Radial Velocity, Transits, Microlensing, etc).  This data provides distributions of planetary parameters that can be sampled from: the planetary mass function (PMF), the distribution of planetary orbital radii, and the host star metallicity distribution. \\

\begin{figure}
\begin{center}
\includegraphics[scale = 0.5]{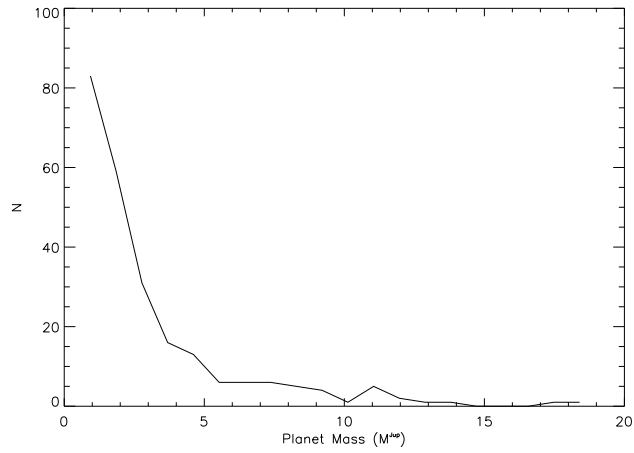}
\caption{\emph{The planetary mass function, constructed from the Exoplanet Encyclopedia data (http://exoplanet.eu)} \label{fig:PIMF}}
\end{center}
\end{figure}

\begin{figure}
\begin{center}
\includegraphics[scale = 0.5]{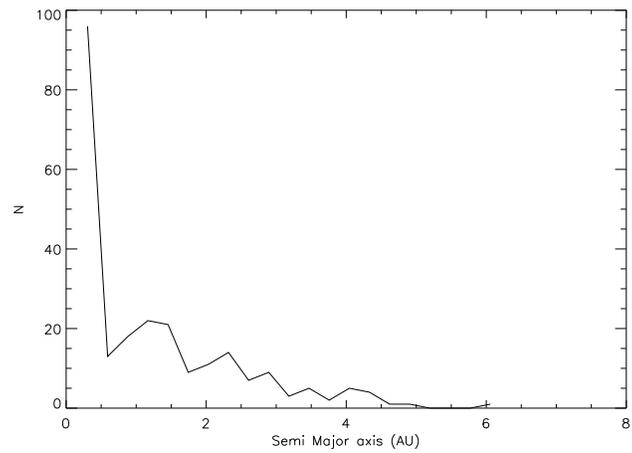}
\caption{\emph{The distribution of planetary orbital radii, constructed from the Exoplanet Encyclopedia data (http://exoplanet.eu)} \label{fig:radii}}
\end{center}
\end{figure}

\noindent  Therefore, as with the stellar parameters, a population of planets can be created around the parent stars, with statistical properties matching what can be observed.  However, this statistical data is still subject to strong observational bias, and the catalogues are still strongly incomplete.  There is insufficient data to reproduce a distribution of terrestrial planets: therefore it is assumed that life evolves around the satellites of the planets simulated here.  In essence, this constitutes a lower limit on the number of inhabited planets: the work of Ida and Lin (2004)\cite{Ida_Lin_metal} shows that, as a function of metallicity, habitable terrestrial planets are comparable in frequency (or higher) than currently detectable giant planets.  This data is hence still useful for illustrating the efficacy of the Monte Carlo method (at least, until observations of terrestrial exoplanets become statistically viable).  All this should be borne in mind when the results of this work are considered. \\

\noindent Two further parameters to be sampled, for which observational data is difficult to obtain, are \(P_{life}\), the probability that life can originate on a given planet, and \(N_{resets}\), the number of biologically damaging events that a planet will experience during the evolution of its indigenous life (such as local supernovae, gamma ray bursts (Annis 1999), cometary impacts, etc).  At this point, the model goes beyond relatively well-constrained parameters, and becomes hypothesis.  \(N_{resets}\) is defined as a function of galactocentric radius -  more resetting events (most notably SNe) occur as distance to the Galactic Centre decreases:

\begin{equation} N_{resets} = \mu_{resets,0}\left(\frac{r_{gal}}{r_{gal,\odot}}\right)^{-1} \end{equation}

\noindent Where \(\mu_{resets,0}\) is set to 5, reflecting the so-called ``Big Five'' mass extinction events in the Earth's fossil record (Raup and Sepkoski (1982)).  It should be noted that this is a simple parameterisation: the real distribution of resetting events is more complex, in particular for short GRBs and Type I SNe, which have less well-defined spatial relationships.  This simple expression is provided to prevent a blurring of the issue: future work will include a more rigorous expression.  All planetary data was collected from the Exoplanets Encyclopaedia website (http://exoplanet.eu): in total, 242 planets were used to construct the data, as these had entries in all required data fields.

\subsection{Biological Parameters}

\subsubsection{Life Parameters}

\noindent The model now enters the realm of essentially pure conjecture: all the available data for these parameters is derived from observations of a single biosphere, and hence there is little that can be done to constrain these parameters (at least without making wide-ranging assumptions about the mechanisms of life as a whole).  The model implicitly assumes the ``hard step'' scenario of evolution (e.g. Carter 2007\cite{stages}), i.e. life must achieve essential evolutionary goals in order to become intelligent organisms with the ability to construct sufficiently complex technological artifacts.  The key biological parameters are listed as follows:

\begin{itemize}
\item \(N_{stages}\): The number of stages life must evolve through to become intelligent
\item \(\tau_i\): The time required for each of these stages to be reached
\item \(\tau_{int}\): The time required for an intelligent technological civilisation to form from life's creation
\item \(P_{annihilate}\): The probability that a reset event will cause complete annihilation 
\end{itemize}

\noindent The intelligence timescale \(\tau_{int}\) is calculated by the following stochastic process: if life does evolve on a planet, \(N_{stages}\) is randomly sampled.  The resetting events \(N_{resets}\) are placed uniformly throughout each of the stages. If \(N_{resets} > N_{stages}\), then any given stage may suffer several reset events. Then the following procedure occurs for each stage \(i\):

\begin{enumerate}
\item \(\tau_i\) is sampled
\item If a reset occurs, test if that reset results in annihilation: if annihilation occurs, life is exterminated, and the process ends; otherwise, \(i\) decreases by 1.
\item \(\tau_{int}\) is increased by some fraction of \(\tau_i\) (or simply by \(\tau_i\) if no resets occur)
\item Increase \(i\) by 1, and return to 1
\end{enumerate} 

\noindent This procedure continues until either a) life has reached the final stage (an intelligent technologically capable civilisation has evolved), or b) the intelligence timescale becomes greater than the main sequence lifetime of the parent star: \(\tau_{int} > \tau_{MS} \).  Once a civilisation has formed, it is assumed that detectable signals or signal leakage begins to be emitted.  This emission will continue until either (a) the civilisation destroys itself or (b) the parent star moves off the main sequence (see next section).  This in itself is a conservative estimate of the length of time a civilisation may be detected, as civilisations may in fact migrate between the stars, or use stellar engineering to prolong their parent's stars life. \\

\noindent The \(P_{annihilate}\) parameter is defined as:

\begin{equation} P_{annihilate} = 1 - e^{r_{gal}-r_{gal,\odot} }\end{equation}

\noindent Along with the metallicity gradient and the \(N_{resets}\) parameter specified in the previous sections, this defines a Galactic Habitable Zone which mimics that defined by Lineweaver et al (2004)\cite{Line_GHZ}: an annular region of the galaxy where the metallicity is sufficiently high for planets (and life) to form, and the number of biologically destructive events is sufficiently small (and their destructive capability sufficiently low) to allow life to evolve at all. \\

\subsubsection{Civilisation Parameters}

\noindent Once a technologically capable civilisation has formed, it must move through a ``fledgling phase'': it is susceptible to some catastrophic event caused partially or fully by its own actions (e.g. war, plague, catastrophic climate change, botched macro-engineering projects).  This is described by the parameter \(P_{destroy}\): the probability that a fledgling civilisation will destroy itself.  If a civilisation can survive this phase, it becomes sufficiently advanced to prevent such self-destruction events, and becomes stable, on a timescale \(\tau_{adv}\).  If a civilisation is destroyed, then it will survive some fraction of \(\tau_{adv}\) before destruction. \\

\noindent What advanced civilisations can then do is at the behest of the user: civilisations may colonise all planets within their solar system, resulting in signals appearing on all planets in that system.  Probes may be sent into the galaxy at large, which could define an explorable volume of the galaxy for a given advanced civilisation.  Civilisations may even attempt to generate new biospheres on neighbouring planets - the ``directed panspermia'' model (Crick and Orgel 1972 \cite{direct_panspermia}).  The results in this paper assume colonisation of all \emph{uninhabited} planets in the system will occur once a civilisation becomes advanced (colonisation being an umbrella term for placing both manned and unmanned installations on the surface or in orbit of said planets)\footnote{It is reasonable to assume that a pre-existing biosphere, perhaps with intelligent life would be a more suitable candidate for colonisation rather than a lifeless planet: however, to prevent statistical confusion (and to avoid dealing with the possibilities of interplanetary conflict between intelligent species) colonisation occurs only on uninhabited planets}. \\

\noindent The key civilisation parameters are:

\begin{itemize}
\item \(\tau_{adv}\): The timescale for a civilisation to move from ``fledgling'' to ``advanced''.
\item \(P_{destroy}\): The probability that a fledgling civilisation will destroy itself.
\item \(L_{signal}\): The lifetime of any signal or leakage from a civilisation 
\end{itemize}

\noindent The signal lifetime of a self-destroying civilisation is:

\begin{equation} L_{signal} = x\tau_{adv} \end{equation}

\noindent Where \(x\) is a uniformly sampled number between 0 and 1.  If the civilisation becomes advanced, this becomes

\begin{equation} L_{signal} = \tau_{MS} - \tau_{int} \end{equation}

\noindent  (i.e. civilisations exist until their parent star leaves the Main Sequence).  For a planet colonised by an advanced civilisation, this is

\begin{equation} L_{signal} = \tau_{MS} - \tau_{int} - \tau_{adv} \end{equation}

\noindent At the end of any MCR run, each planet will have been assigned a habitation index based on its biological history.

\begin{equation} I_{inhabit} = \left\{
\begin{array}{l l }
-1 & \quad \mbox{Biosphere which has been annihilated} \\
0 & \quad \mbox{Planet is lifeless} \\
1 & \quad \mbox{Planet has primitive life} \\
2 & \quad \mbox{Planet has intelligent life} \\
3 & \quad \mbox{Planet had intelligent life, but it destroyed itself} \\
4 & \quad \mbox{Planet has an advanced civilisation} \\
5 & \quad \mbox{Planet has been colonised by an advanced civilisation} \\
\end{array} \right. \end{equation}

\noindent If the planet has habitation index 1 or 2, the biological process has been ended by the destruction of the parent star.  Planets with an index of -1 or 1 may contain biomarkers in their atmosphere (e.g. ozone or water spectral features) which could be detected.  Planets with an index of 2 or higher will emit signals or signal leakage.  Planets with an index of 4 or 5 may display evidence of a postbiological civilisation or of large scale ``macro-engineering'' projects , e.g. Dyson spheres (Dyson 1960\cite{Dyson}).  Signals from these systems may even be consistent with those expected from  Kardashev Type II civilisations (those which can harness all the energy of their parent star), and hence could produce characteristic stellar spectral signatures that Earth astronomers could detect.

\section{Inputs}\label{sec:Inputs}

\noindent When comparing hypotheses, it is important to keep the number of free parameters to a minimum.  To achieve this, a biological version of the Copernican Principle is invoked: the Terran biosphere is not special or unique.  Therefore, life on other planets will share similar values of characteristic parameters.  This implies that many parameters can be held constant throughout all hypotheses: in particular, those that deal with the approach to intelligence.  These are sampled from Gaussian distribution functions (assuming that each variable is the result of many subfactors, and implicitly applying the Central Limit Theorem).  The mean value for \(N_{stages}\) is taken to be 6, reflecting the major stages life went through on Earth - biogenesis, the advent of bacteria, the advent of eukaryotes, combigenesis, the advent of metazoans, and the birth of technological civilisation (Carter 2007 \cite{stages}).  100 MCRs were produced for each hypothesis: the parameters used for all testing can be found in \textbf{Table \ref{tab:common}}.  The three hypotheses tested are described below.

\subsection{The Panspermia Hypothesis}

\noindent This well-documented theory suggests life may spread from one originating planet to many others, causing life to form concurrently in multiple systems.  In this model, if life forms on any planet in a star system, then other planets may be seeded, according to the following prescription:

\begin{equation} P_{life} = \left\{
\begin{array}{l l }
1 & \quad \mbox{if planet is in stellar habitable zone} \\
e^{-\Delta R} e^{-\left(\frac{L_*}{L_{\odot}}\right)}  & \quad \mbox{if any planet in the same solar system is inhabited} \\
\end{array} \right. \label{eq:panspermia} \end{equation}

\noindent Life can move between planets if the distance between the destination planet and the original inhabited planet (\(\Delta R\)) is sufficiently small, and the stellar luminosity is sufficiently low that ionising radiation will not destroy biological organisms in transit.  It is assumed that life emerges on the original planet while a significant amount of planetary bombardment is in progress, and the seeding of other planets (from rocky fragments expelled from the inhabited planet's surface by asteroid impacts) is therefore effectively instantaneous. \(P_{destroy}\) is specified \emph{a priori} to be 0.5, reflecting current ignorance about the development of extraterrestrial civilisation. \\

\noindent This hypothesis restricts panspermia to planets in the same star system only: unfortunately, the model is currently incapable of modelling the interesting possibility of interstellar panspermia (Napier 2004 \cite{Napier_IP}, Wallis and Wickramasinghe 2004 \cite{Wallis_IP}).  This in itself is strong motivation to improve the model further, and is left as an avenue for future work.

\subsection{The Rare Life Hypothesis}

\noindent Due to the current dearth of data on Earth-like exoplanets, the famous ``Rare Earth Hypothesis'' (Ward and Brownlee (2000)\cite{rare_Earth}) cannot be tested fully.  A different hypothesis is possible (assuming Jovian planets in the habitable zone may have stable Earth-like moons on which life could arise) with this code. The hypothesis is as follows: once life arises, it is assumed to be tenacious, and can easily evolve into intelligence: however, the initial appearance of life is itself hard: the planetary niche must satisfy certain criteria before life can evolve. The criteria are:

\begin{enumerate}
\item The planet is in the stellar habitable zone
\item The parent star mass is less than \(2M_{\odot}\)
\item The system has at least one extra planet (protection from bombardment)
\item The star's metallicity must be at least solar or higher (\(Z_*\geq Z_{\odot}\))
\end{enumerate}

\noindent If \emph{all} criteria are met, then life can evolve.  \(P_{destroy}\) is set to 0.5 for the same reason as previously.  It should be noted again that this is significantly different from the Rare Earth hypothesis, and should not be considered as a comparison.  A more accurate comparison would have the last few stages of life (the development of metazoans and intelligent life) be rare: this would be better suited for future work after the model has been refined.

\subsection{The Tortoise and Hare Hypothesis}

\noindent As a final example, this hypothesis demonstrates the ability of the code to model more sociological hypotheses.  Life evolves easily on many planets, but the evolution towards intelligence and advanced civilisation is more complex.  Essentially, civilisations which arise too quickly are more susceptible to self-destruction, whereas those which take longer to emerge are more likely to survive the fledgling phase.  This is parametrised as:

\begin{equation} P_{destroy} = \frac{\tau_0}{\tau_{int}} \label{eq:hare}\end{equation}

\noindent Where \(\tau_0 = N_{stages}\tau_{min}\) is a normalisation constant which describes the minimum evolution timescale (if \(N_{resets}=0\), and each stage takes the minimum amount of time \(\tau_{min}\)) (Note:If this calculation is applied to Mankind, our own destruction probability is around 0.8 (!)).  If a planet is in the habitable zone, then \(P_{life} = 1\), in the same vein as the other hypotheses.  It should be noted that this is an oversimplification: the potential existence of liquid water on Enceladus and Europa (both outside the solar habitable zone) suggests that \(P_{life}\) should be a distribution in orbital radius with tails that extend outside the zone: however, for the purposes of this paper, a simple step function will be sufficient.

\begin{table}[h]
 \centering
  \caption{Parameters used for all hypotheses \label{tab:common}}
  \begin{tabular}{l|lcc}
  \hline
  \hline

  \hline
  \hline
   Hypothesis & Parameter &  Mean & Standard Deviation  \\
  \hline
   All & \(N_{stages}\) & 6 & 1 \\
   All & \(\tau_i\) & 0.8 Gyr  & 0.25 Gyr \\
   All & \(\tau_{adv}\) & 2.5 \(\times 10^{-4}\) Gyr & 1.0 \(\times 10^{-4}\) Gyr \\
 \hline
\hline
 Hypothesis & Parameter & Value \\
\hline
 All &  \(N_{MCR}\) & 100 \\
\hline
 Panspermia & \(P_{life}\) &  as equation (\ref{eq:panspermia}) \\
 Panspermia & \(P_{destroy}\) & 0.5 \\
 \hline
 Rare Life & \(P_{life}\) &   1 (if all criteria met)\\
 Rare Life & \(P_{destroy}\) & 0.5 \\
\hline
 Tortoise \& Hare &  \(P_{life}\) &   1  (if in stellar habitable zone)\\
 Tortoise \& Hare &  \(P_{destroy}\) & as equation (\ref{eq:hare}) \\
\hline
\hline
\end{tabular}
\end{table}

\section{Results}\label{sec:Results}

\subsection{The Panspermia Hypothesis}

\noindent The single number statistics for this hypothesis can be seen in \textbf{Table \ref{tab:pan_stats}}. As expected, around half of all emerging civilisations destroy themselves (as \(P_{destroy} = 0.5\)).  Around 0.1\% of all planets are inhabited by either primitive or intelligent life during the Galaxy's lifetime, and roughly 10\% of all inhabited planets host intelligent lifeforms.  Compared to the Tortoise and Hare hypothesis (see below), the inhabited fraction has not greatly increased with possibility of seeding amongst neighbours: this is mostly due to the low multiplicity of current exoplanets (around 11\%).

\begin{table}[h]
\centering
\caption{Statistics for the Panspermia Hypothesis \label{tab:pan_stats}}
\begin{tabular}{lcc}
\hline
\hline
Variable & Mean & Standard Deviation \\
\hline
\(N_{planets,total}\) & \(4.7295 \times 10^8\) & 401530 \\
\(N_{inhabited}\) & 690983.63 & 53 \\
\(N_{fledgling}\) & 75923.04 & 8 \\
\(N_{destroyed}\) & 37958.07 & 11 \\
\(N_{advanced}\) & 37964.97 & 20 \\
\hline
\hline
\end{tabular}
\end{table}

\begin{figure}
$
\begin{array}{cc}
\includegraphics[scale = 0.4]{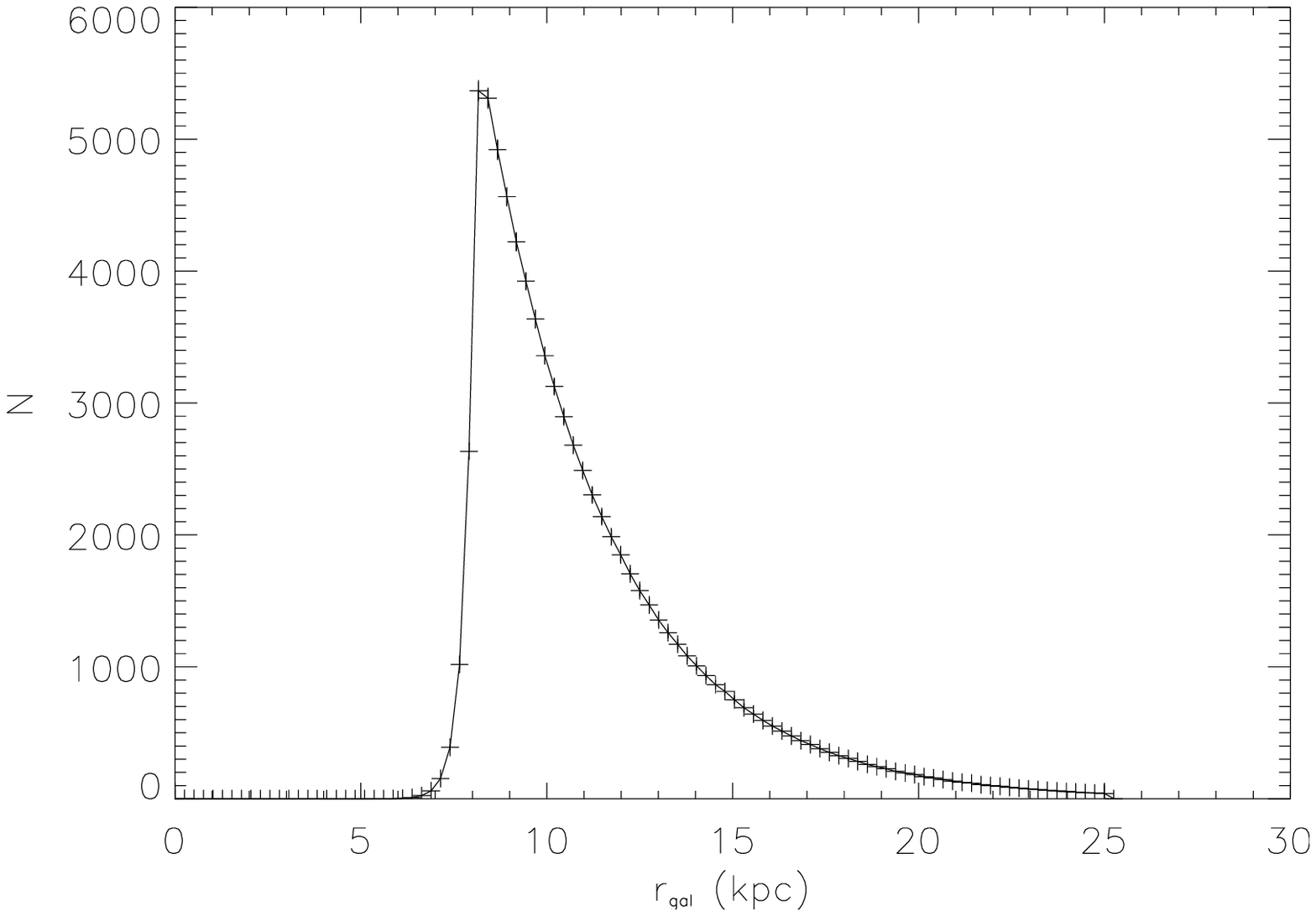} &
\includegraphics[scale = 0.4]{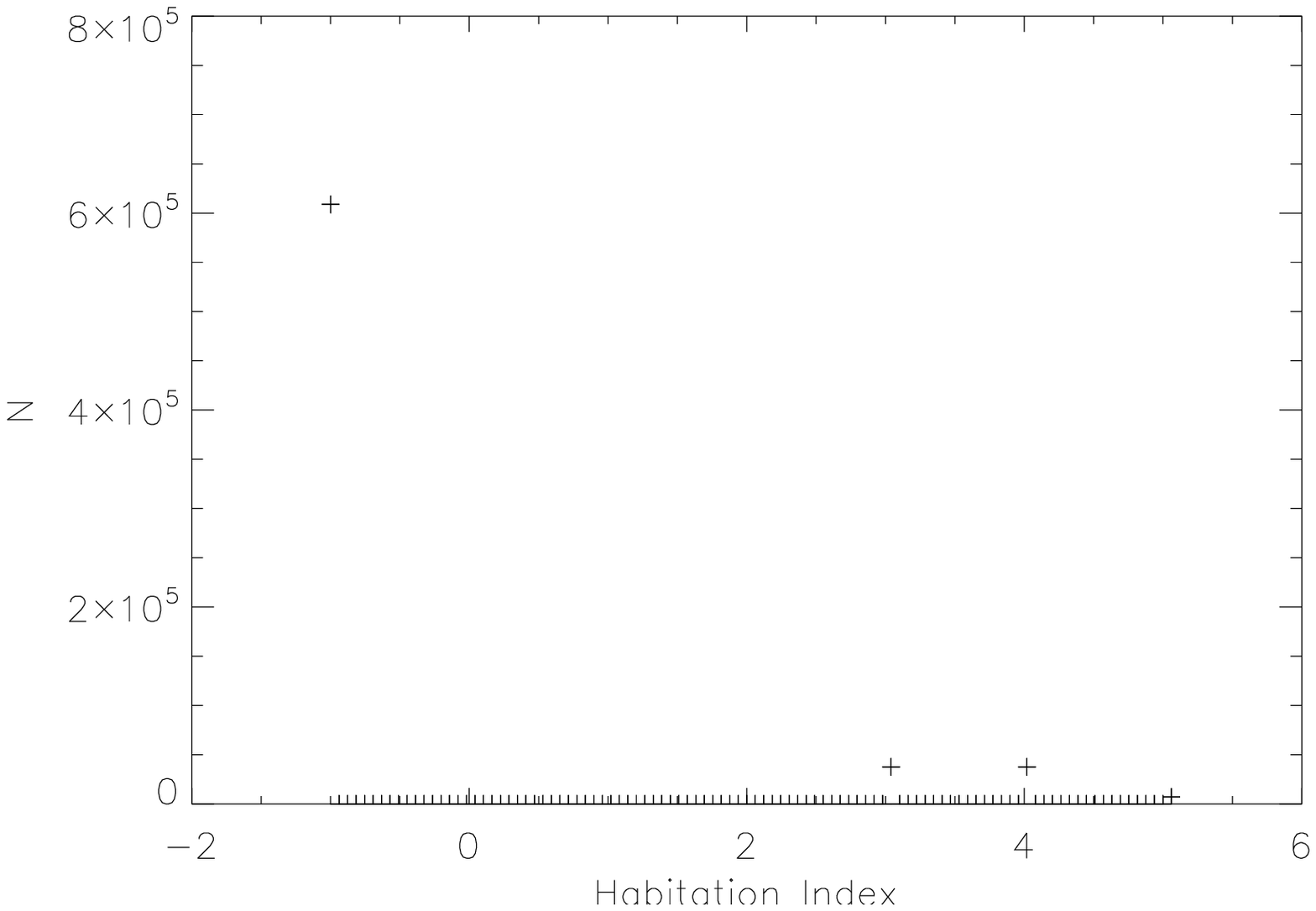} \\
\end{array}$
\caption{\emph{The distribution of galactocentric radius (left) and habitation index (right) under the Panspermia Hypothesis.}\label{fig:rhisto_pan}}
\end{figure}

\noindent \textbf{Figure \ref{fig:rhisto_pan}} shows the Galactic Habitable Zone for this hypothesis: at low galactocentric radii, the annihilation rate increases to the point where civilisations cannot develop (although life itself can emerge there, it is quickly eradicated by destructive astrophysical events).  The right panel of \textbf{Figure \ref{fig:rhisto_pan}} shows the breakdown of inhabited planets by type over the life of the Galaxy.  The large number of planets with index -1 (annihilated biospheres) indicates the solar systems with low galactocentric radii which are metal rich, and form planets (and hence life) more easily, but suffer rapid biological catastrophe.  The colonisation fraction (planets with habitation index 5) is quite low: this is most likely due to panspermia seeding reducing the number of neighbouring empty planets available for colonisation. \\

\begin{figure}
$
\begin{array}{cc}
\includegraphics[scale = 0.4]{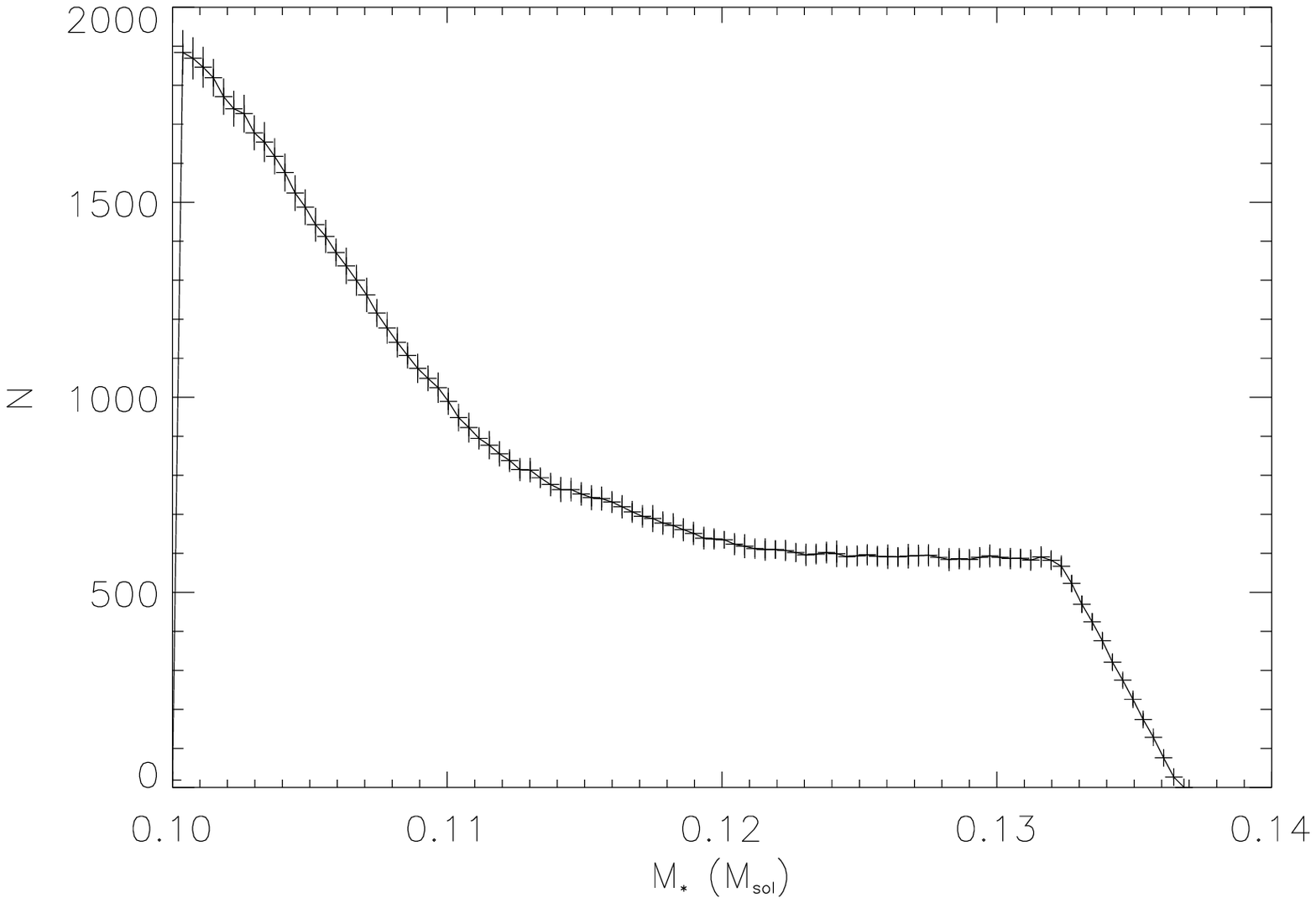} &
\includegraphics[scale = 0.4]{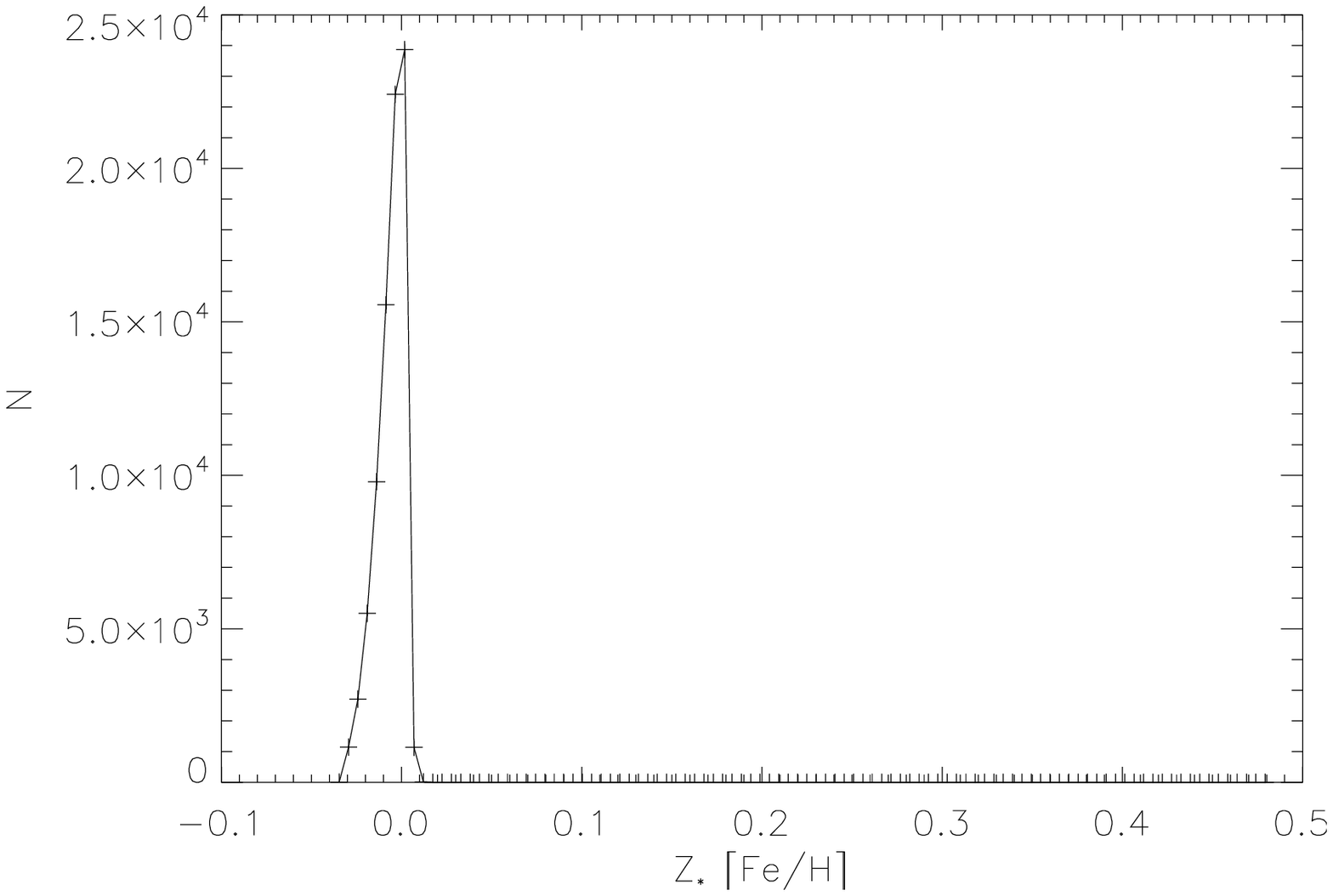} \\
\end{array}$
\caption{\emph{The distribution of host star mass (left) and metallicity (right) under the Panspermia Hypothesis.} \label{fig:mstar_pan}}
\end{figure}

\begin{figure}
$
\begin{array}{cc}
\includegraphics[scale = 0.4]{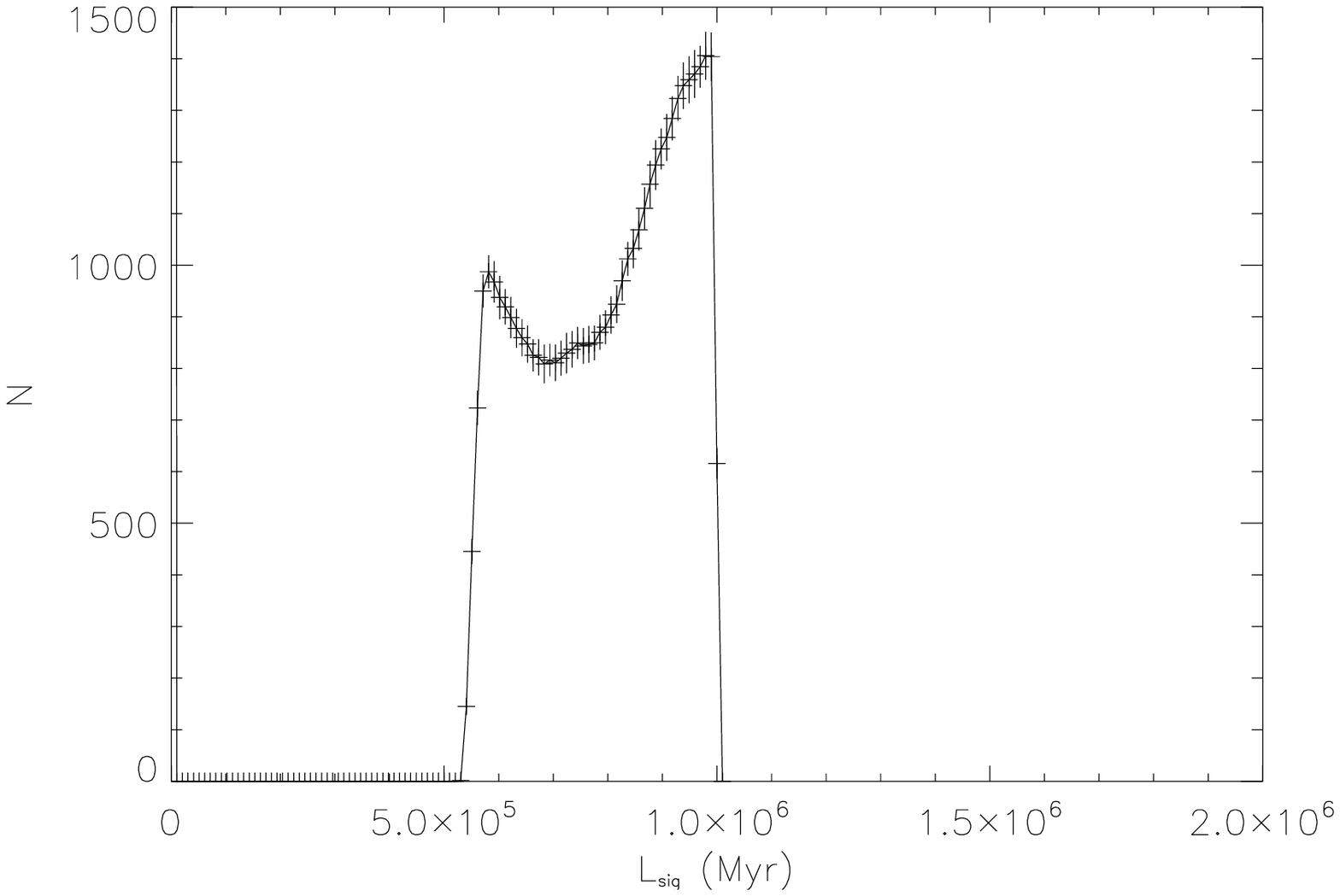} &
\includegraphics[scale = 0.4]{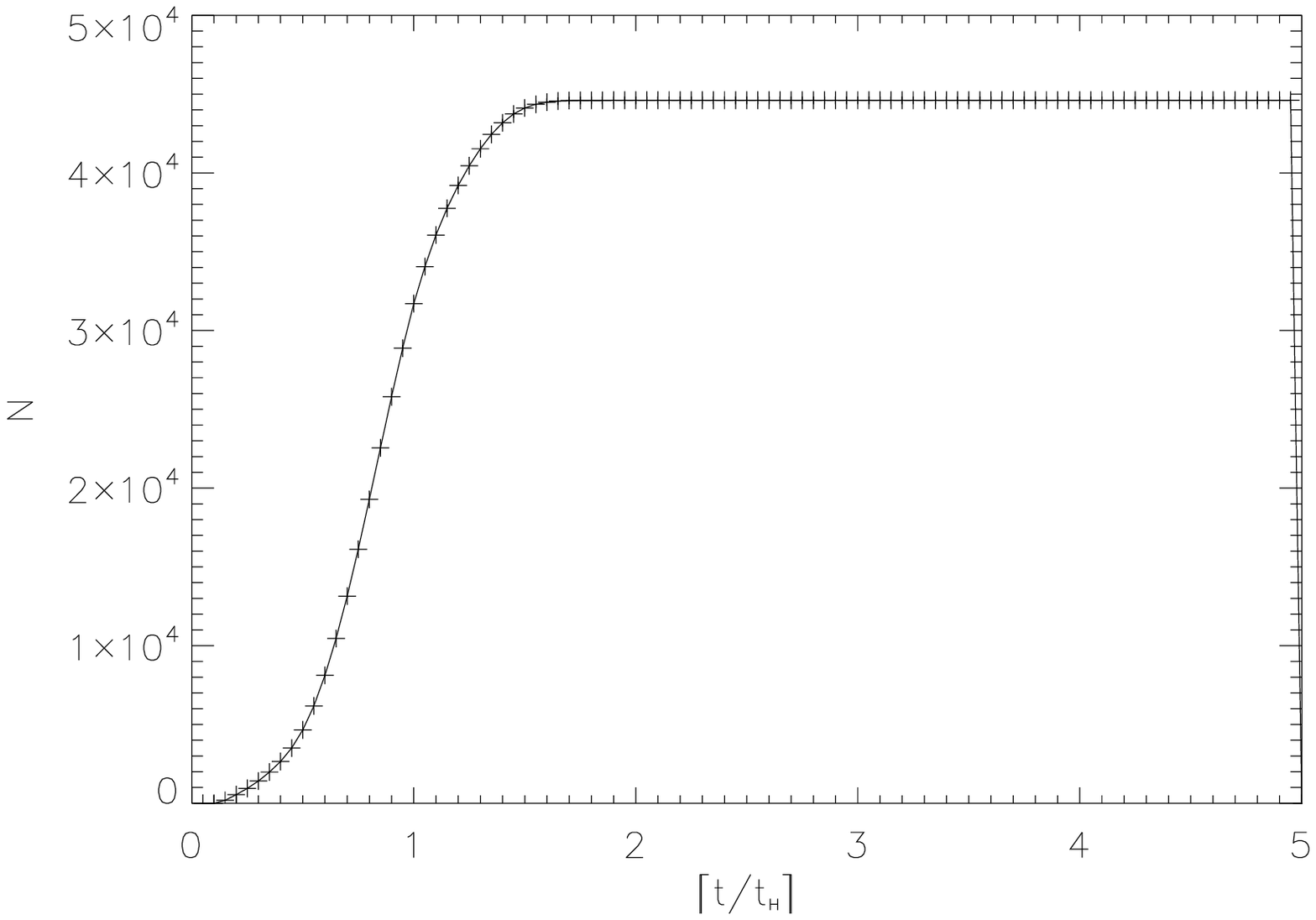} \\
\end{array}$
\caption{\emph{The distribution of signal lifetimes (left) and the signal history of the Galaxy (right) under the Panspermia Hypothesis.} \label{fig:Lsig_pan}}
\end{figure}

\noindent The distribution of stellar properties for civilised planets (\textbf{Figure \ref{fig:mstar_pan}}) indicates a predilection for low mass stars.  This comes from the selection bias of the so-called ``Hot Jupiters'': these objects are of such low orbital radii that if they are to exist in the habitable zone, it must also be of low radius, and hence belong to a low-mass star.  The metallicity reflects the Galactic Habitable Zone: civilisations have a minimum galactocentric radius, and hence a maximum metallicity.  Considering the signal lifetime distribution, it is heavily influenced by the main sequence lifetime of the parent stars, which is a function of mass.  Studying the mass relation, it is clear that the two are correlated.  If the signals are plotted as a function of time (\textbf{Figure \ref{fig:Lsig_pan}}, right panel), it can be seen that the signal history undergoes a transition from low to high N around \(t \sim t_H\), where \(t_H\) is the Hubble time (i.e., the age of the universe).  This is symptomatic of the Copernican Principle invoked to constrain the majority of the biological parameters: on average, any given civilisation will have a intelligence timescale similar to ours.

\subsection{The Rare Life Hypothesis}

\noindent The stringent nature of this hypothesis can be seen in its decreased population of life-bearing planets in comparison to the other hypotheses (with an inhabited fraction of only 0.01\%).  However, despite this stringency, there is a significant number of civilisations forming over the Galaxy's lifetime: around 1\% of all life-bearing planets produce intelligent species.  Again, as expected around half of all civilisations destroy themselves.

\begin{table}[h]
\centering
\caption{Statistics for the Rare Life Hypothesis \label{tab:rare_stats}}
\begin{tabular}{lcc}
\hline
\hline
Variable & Mean & Standard Deviation \\
\hline
\(N_{planets,total}\) & \(4.770 \times 10^8\) & 381 \\
\(N_{inhabited}\) & 80090 & 22 \\
\(N_{fledgling}\) & 728.6 & 1 \\
\(N_{destroyed}\) & 367.3 & 0 \\
\(N_{advanced}\) & 361.2 & 2 \\
\hline
\hline
\end{tabular}
\end{table}

\begin{figure}
$
\begin{array}{cc}
\includegraphics[scale = 0.4]{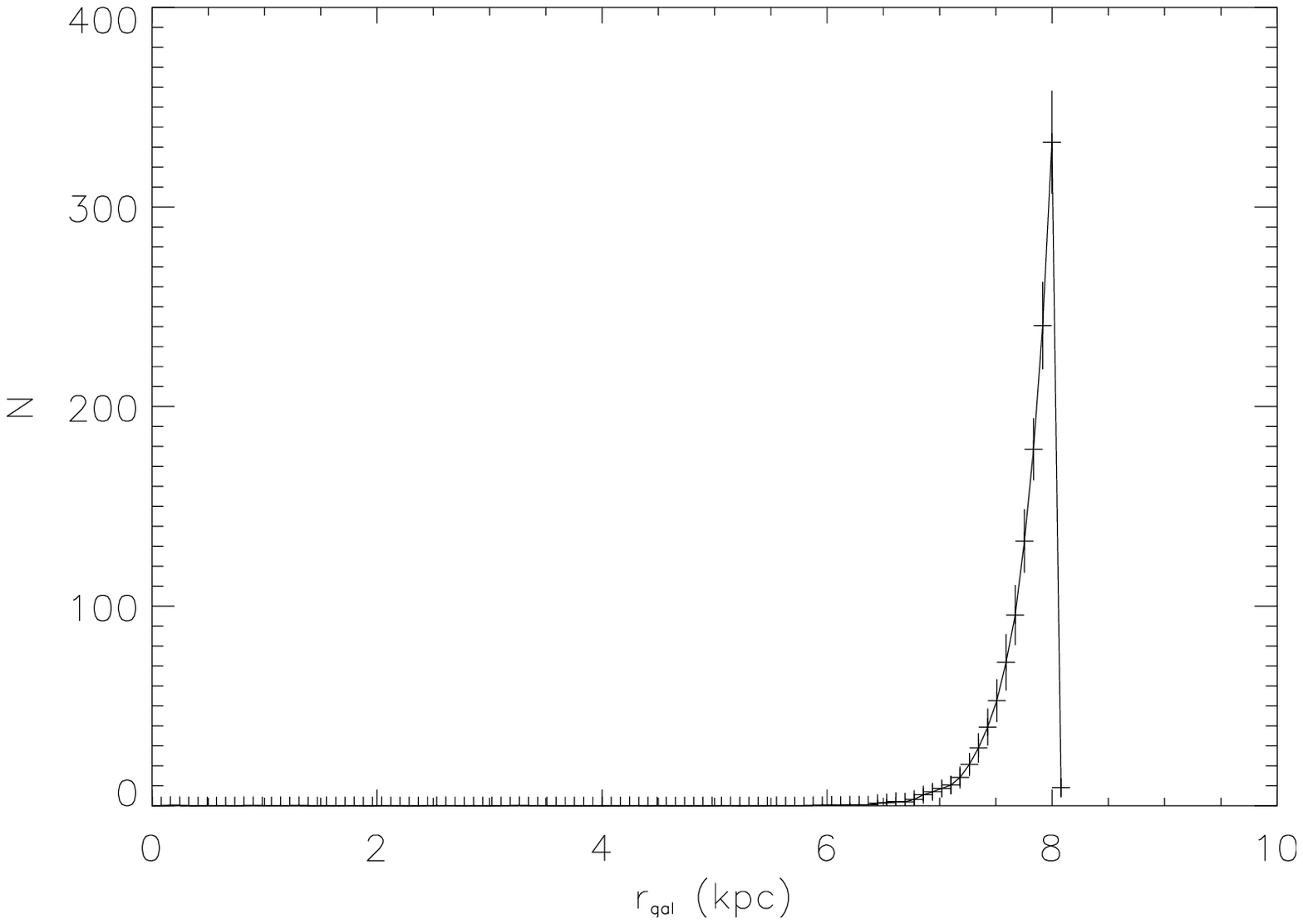} &
\includegraphics[scale = 0.4]{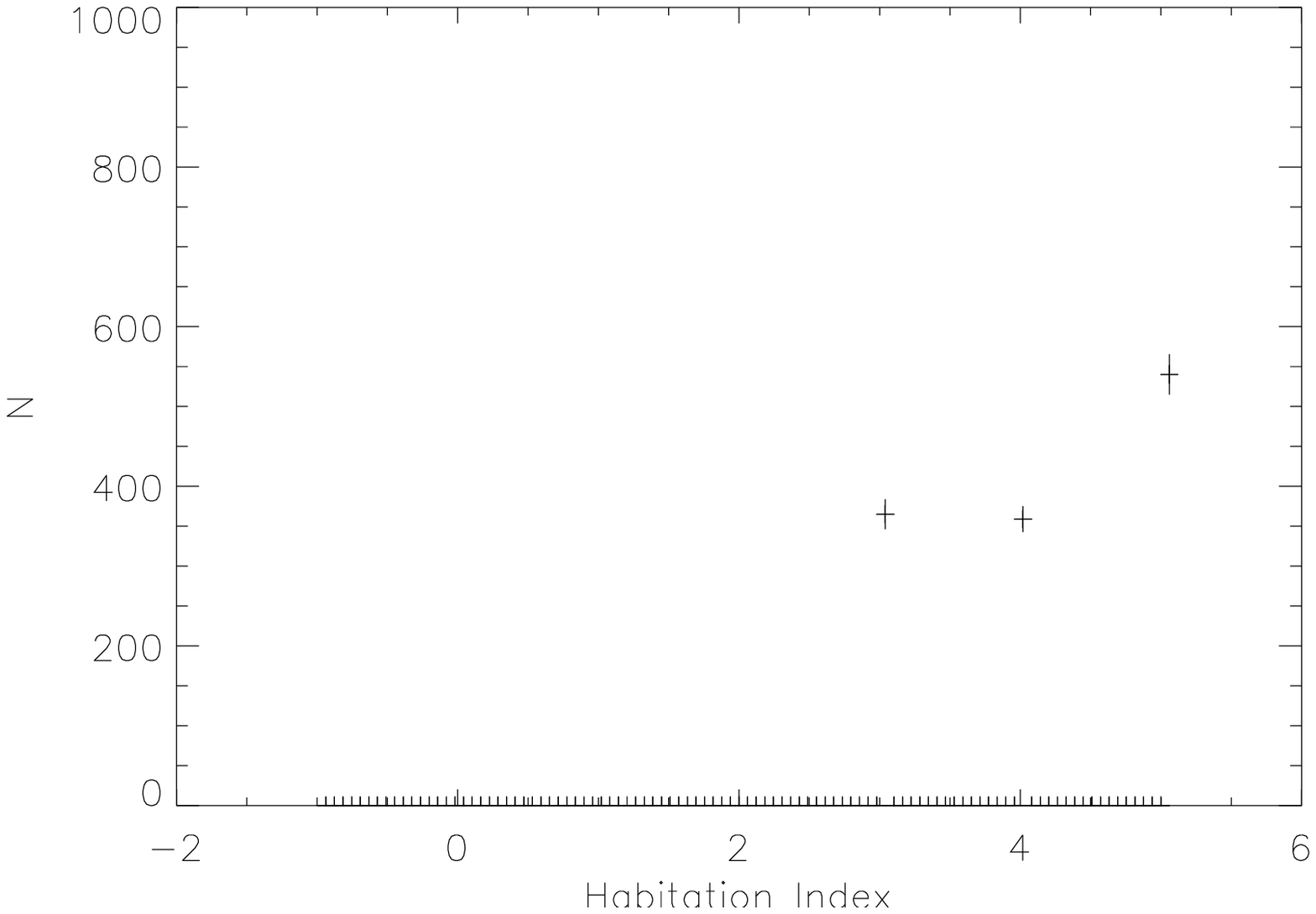} \\
\end{array}$
\caption{\emph{The distribution of galactocentric radius (left) and habitation index (right) under the Rare Life Hypothesis.}\label{fig:rhisto_rare}}
\end{figure}

\begin{figure}
$
\begin{array}{cc}
\includegraphics[scale = 0.4]{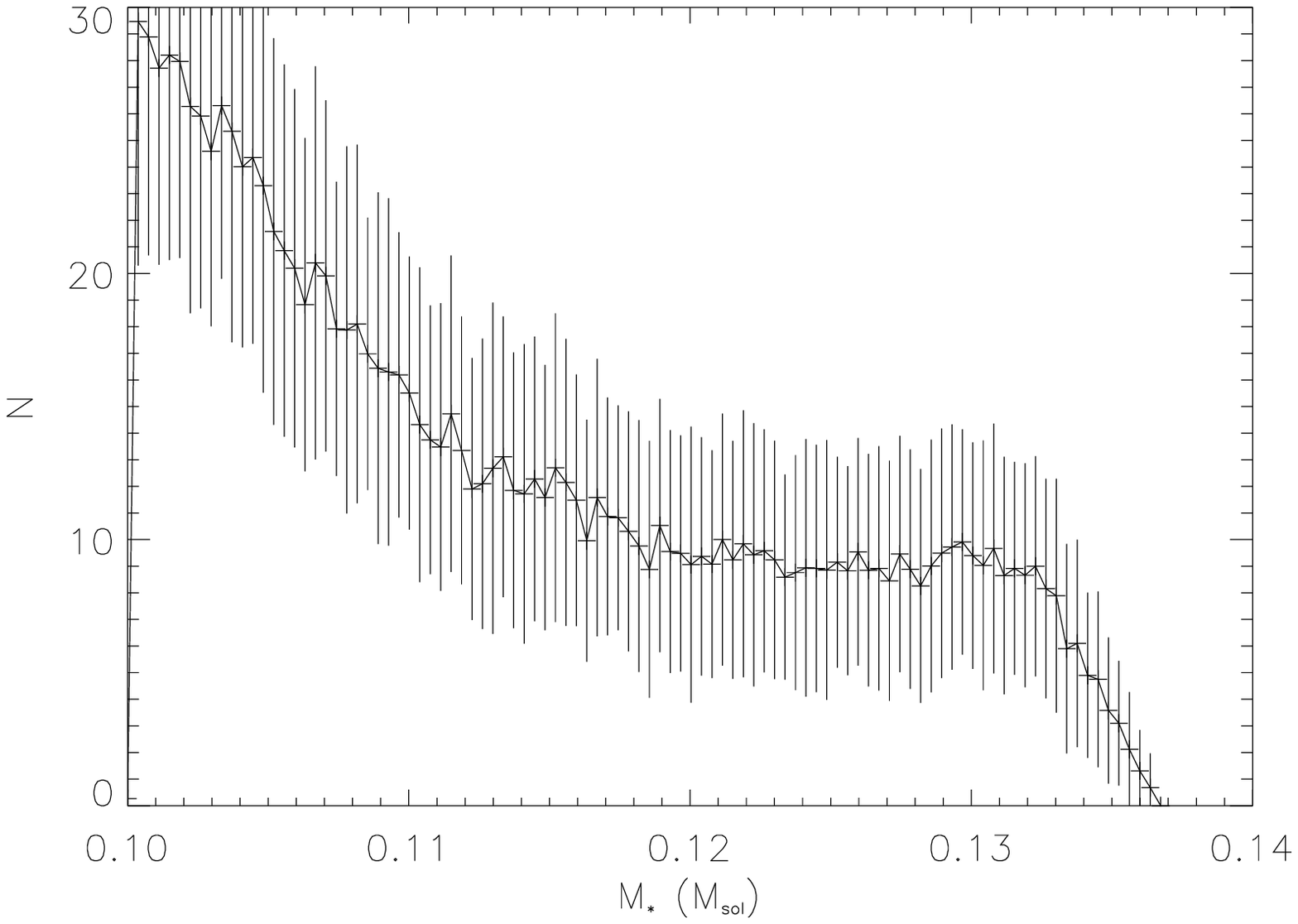} &
\includegraphics[scale = 0.4]{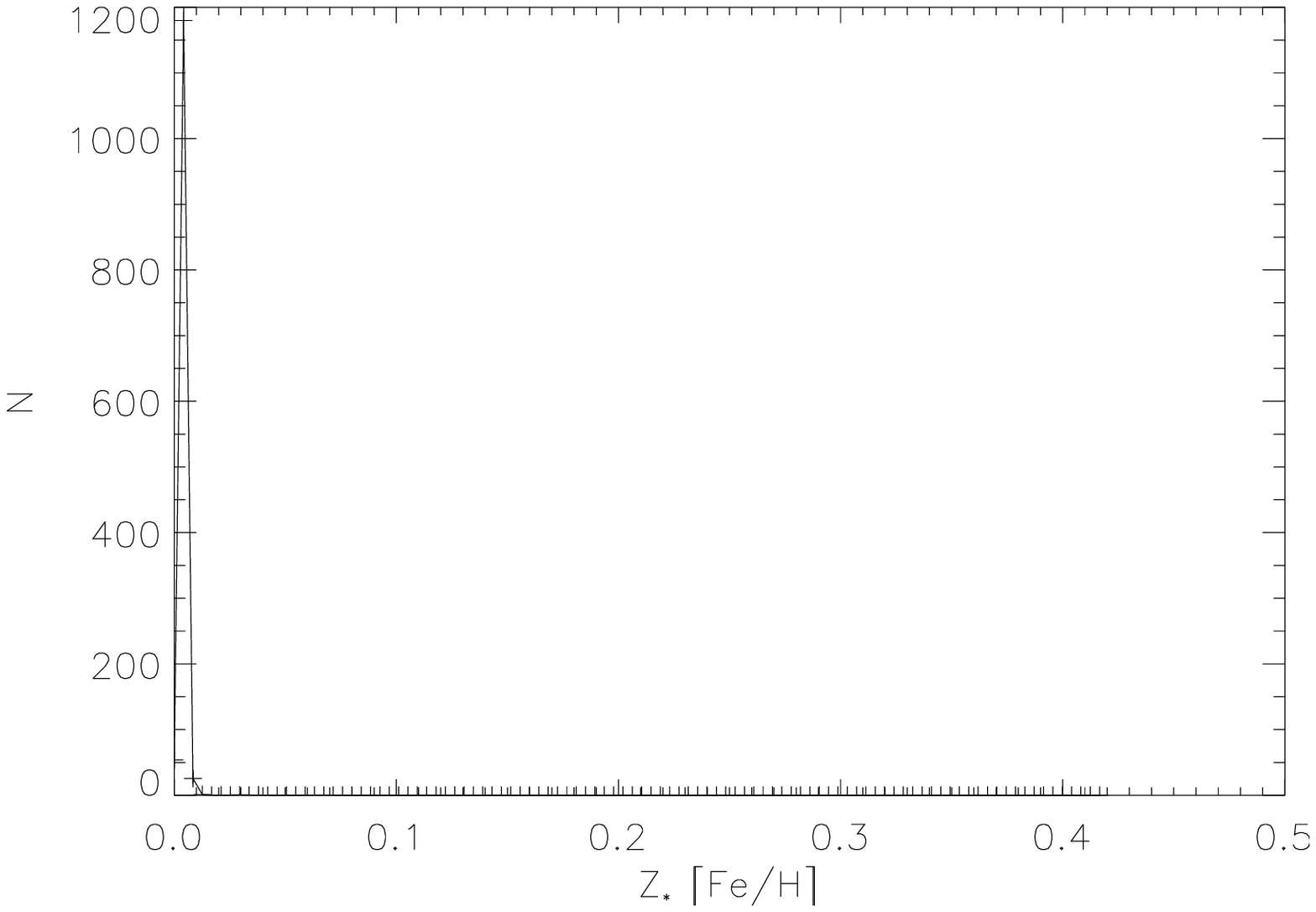} \\
\end{array}$
\caption{\emph{The distribution of host star mass (left) and metallicity (right) under the Rare Life Hypothesis.} \label{fig:mstar_rare}}
\end{figure}

\noindent The distribution of galactocentric radii (\textbf{Figure \ref{fig:rhisto_rare}}, left panel) is quite sharply peaked: this is because life requires \(Z_*\geq Z_{\odot}\) (which is clearly seen in \textbf{Figure \ref{fig:mstar_rare}}, right panel) under this hypothesis, and fixes a maximum \(r_{gal}\), beyond which life cannot begin.   Considering the habitation index (\textbf{Figure \ref{fig:rhisto_rare}}, right panel), it can be seen that there is a great deal of colonisation occurring.  This is due to the criterion that life can only develop in multiple planetary systems, and hence any intelligence which becomes advanced \emph{must} take part in colonisation.  The distribution of star mass, although greatly reduced, is qualitatively the same as for the Panspermia Hypothesis, and hence the signal lifetime has a similar distribution also.

\begin{figure}
$
\begin{array}{cc}
\includegraphics[scale = 0.4]{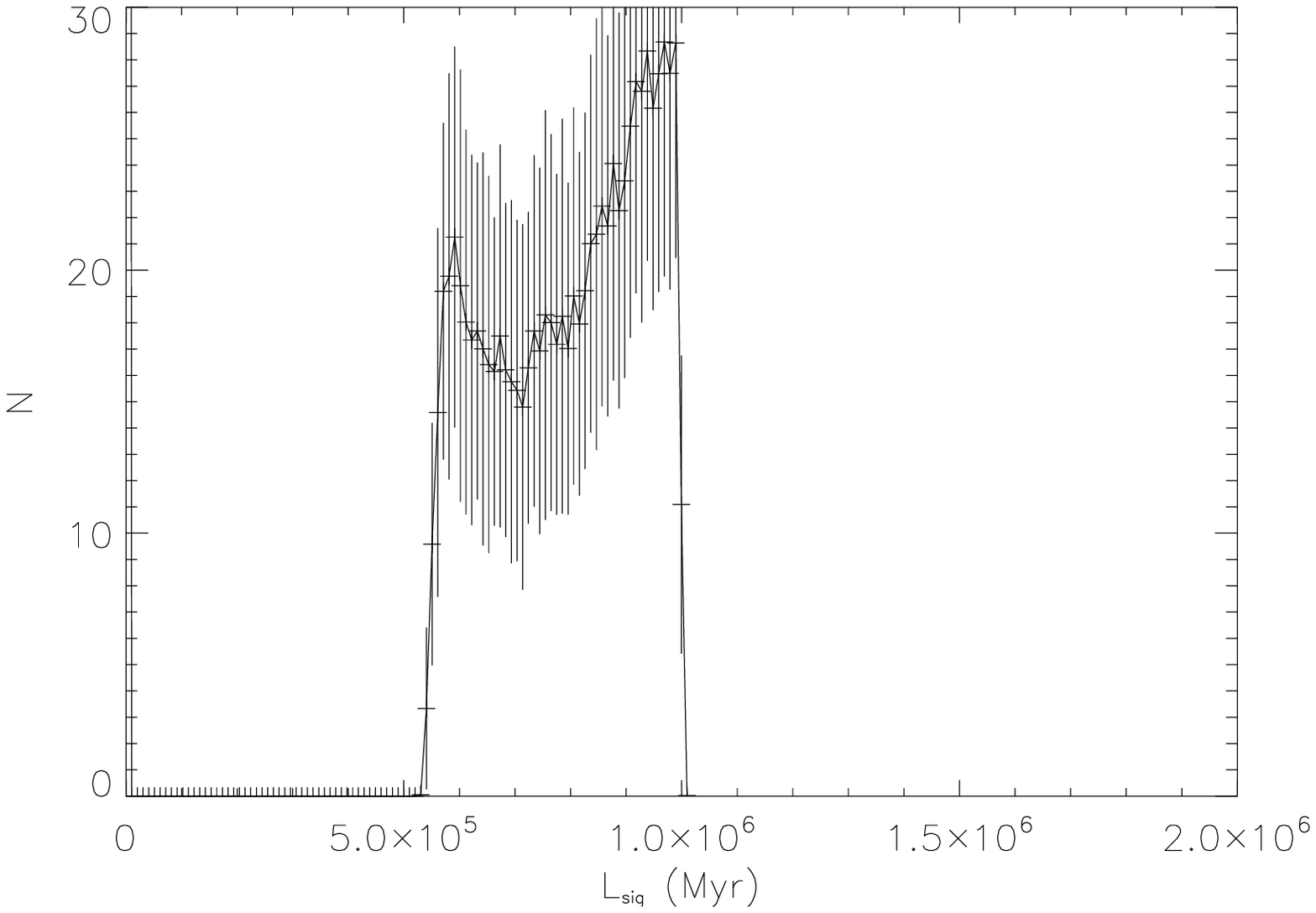} &
\includegraphics[scale = 0.4]{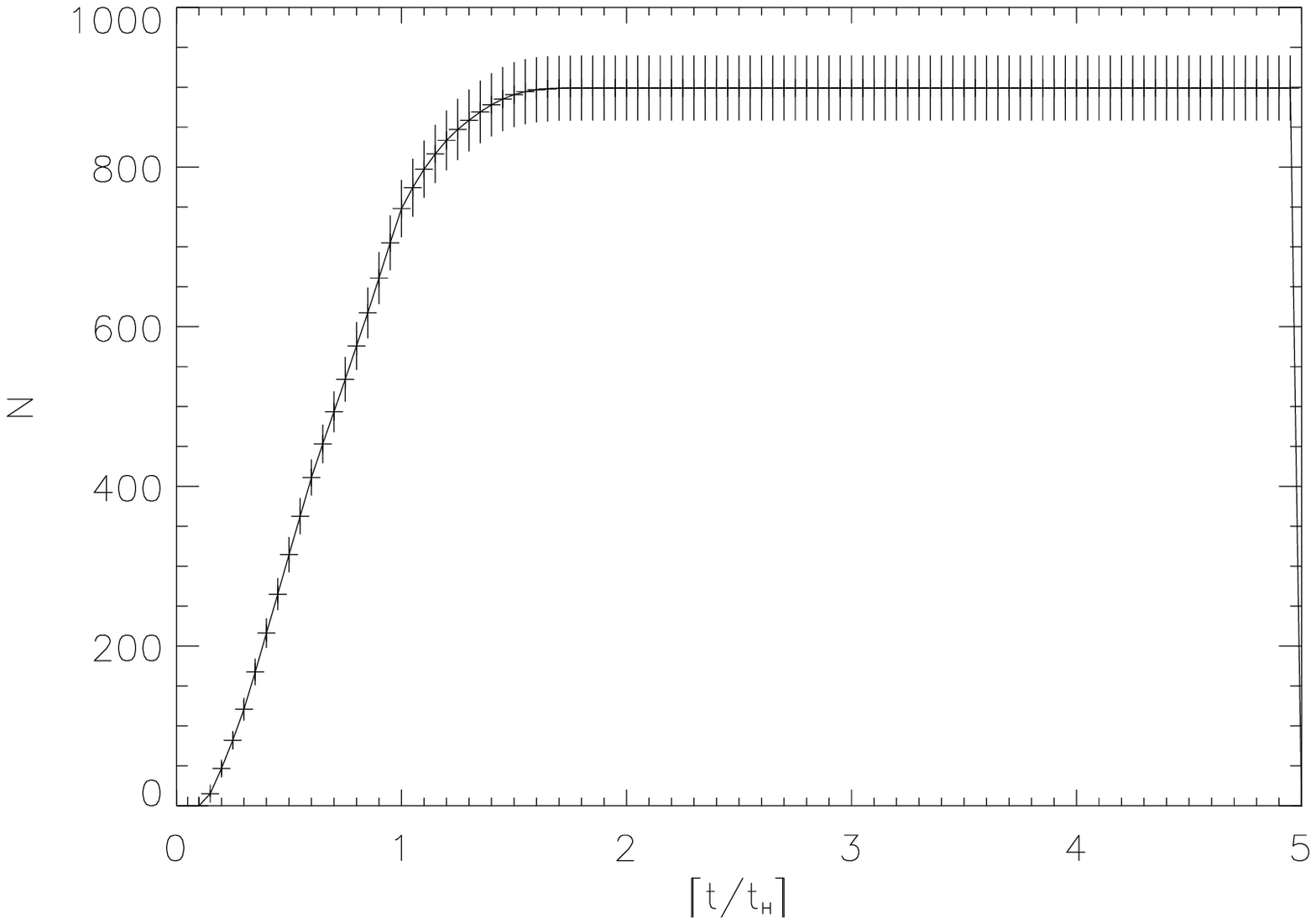} \\
\end{array}$
\caption{\emph{The distribution of signal lifetimes (left) and the signal history of the Galaxy (right) under the Rare Life Hypothesis.} \label{fig:Lsig_rare}}
\end{figure}

\subsection{The Tortoise and Hare Hypothesis}

\noindent The biological data for this hypothesis is not markedly different from the Panspermia Hypothesis, again with an inhabited fraction of around 0.1\%. However, the true difference is in the civilisation data: instead of there being an approximate 1:1 ratio between self-destruction and advancement, it can be seen that this hypothesis weakly favours self-destruction for a given fledgling civilisation, with a ratio of approximately 1.38:1.  Despite this, around 4\% of inhabited planets develop intelligent life that survives the fledgling phase. \\

\begin{table}[h]
\centering
\caption{Statistics for the Tortoise and Hare Hypothesis \label{tab:hare_stats}}
\begin{tabular}{lcc}
\hline
\hline
Variable & Mean & Standard Deviation \\
\hline
\(N_{planets,total}\) & \(4.770 \times 10^8\) & 1593 \\
\(N_{inhabited}\) & 684399.26 & 2 \\
\(N_{fledgling}\) & 75200.3 & 20 \\
\(N_{destroyed}\) & 43626.82 & 1 \\
\(N_{advanced}\) & 31573.52 & 20 \\
\hline
\hline
\end{tabular}
\end{table}

\begin{figure}
$
\begin{array}{cc}
\includegraphics[scale = 0.4]{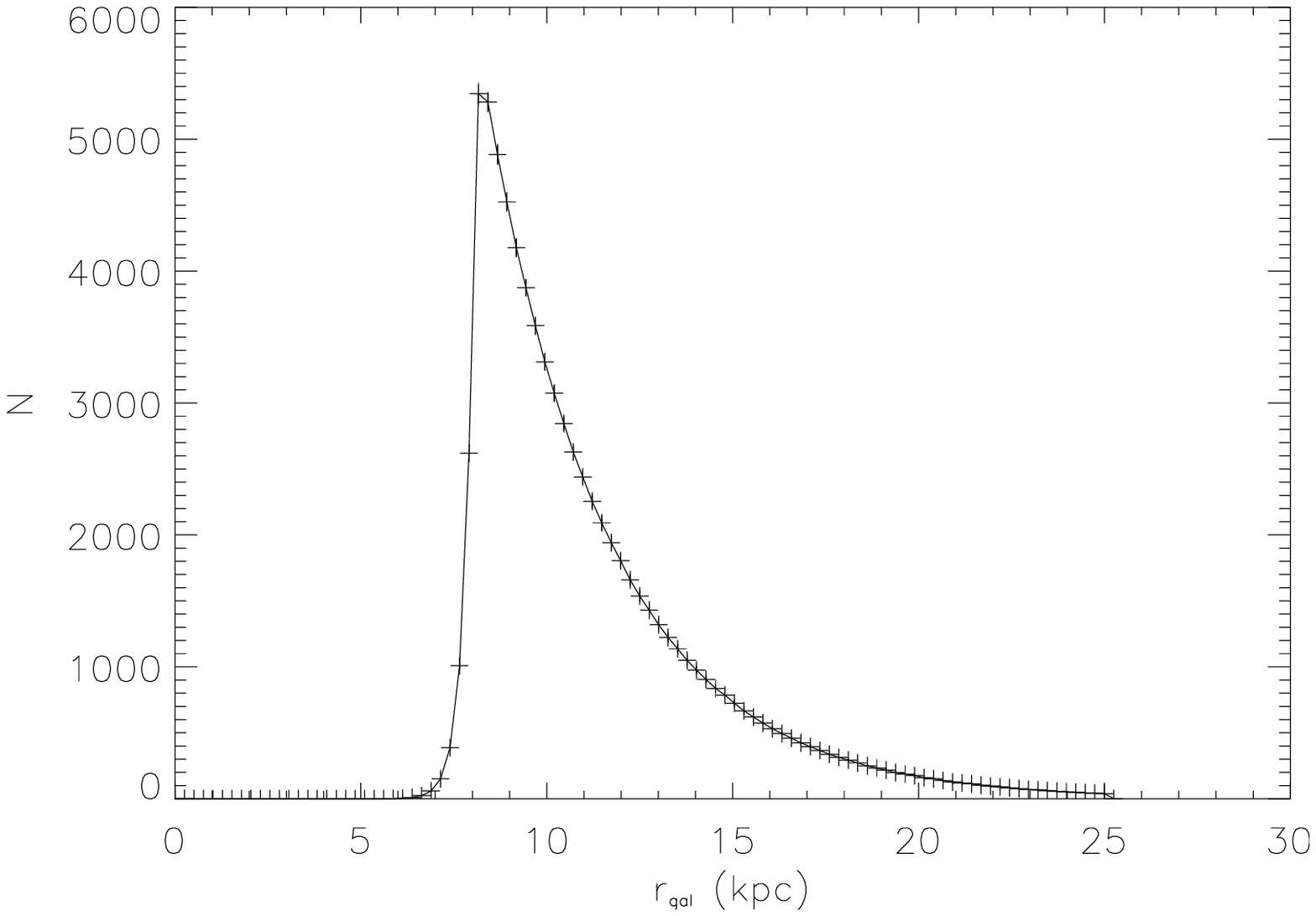} &
\includegraphics[scale = 0.4]{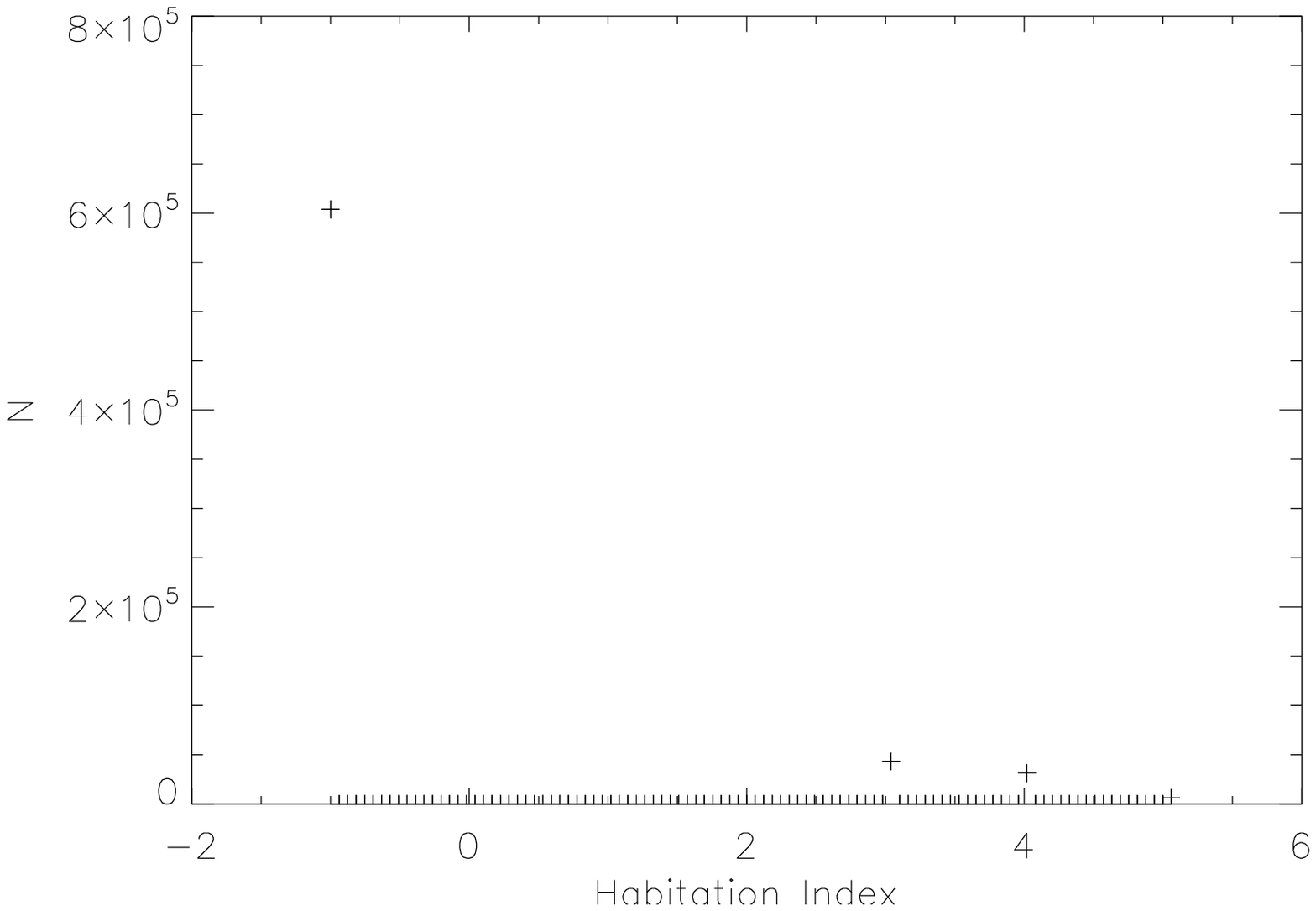} \\
\end{array}$
\caption{\emph{The distribution of galactocentric radius (left) and habitation index (right) under the Tortoise and Hare Hypothesis.}\label{fig:rhisto_hare}}
\end{figure}

\noindent The Galactic Habitable Zone is similar to that for the Panspermia Hypothesis: the weak favouring of self-destruction can be seen in the right panel of \textbf{Figure \ref{fig:rhisto_hare}}.  Comparing the data between this hypothesis and the Panspermia Hypothesis, both the qualitative and the quantitative behaviour is similar.  This reflects the similarity of the two hypotheses in terms of biological parameters, and that the key difference is in the sociological parameter \(P_{destroy}\).

\begin{figure}
$
\begin{array}{cc}
\includegraphics[scale = 0.4]{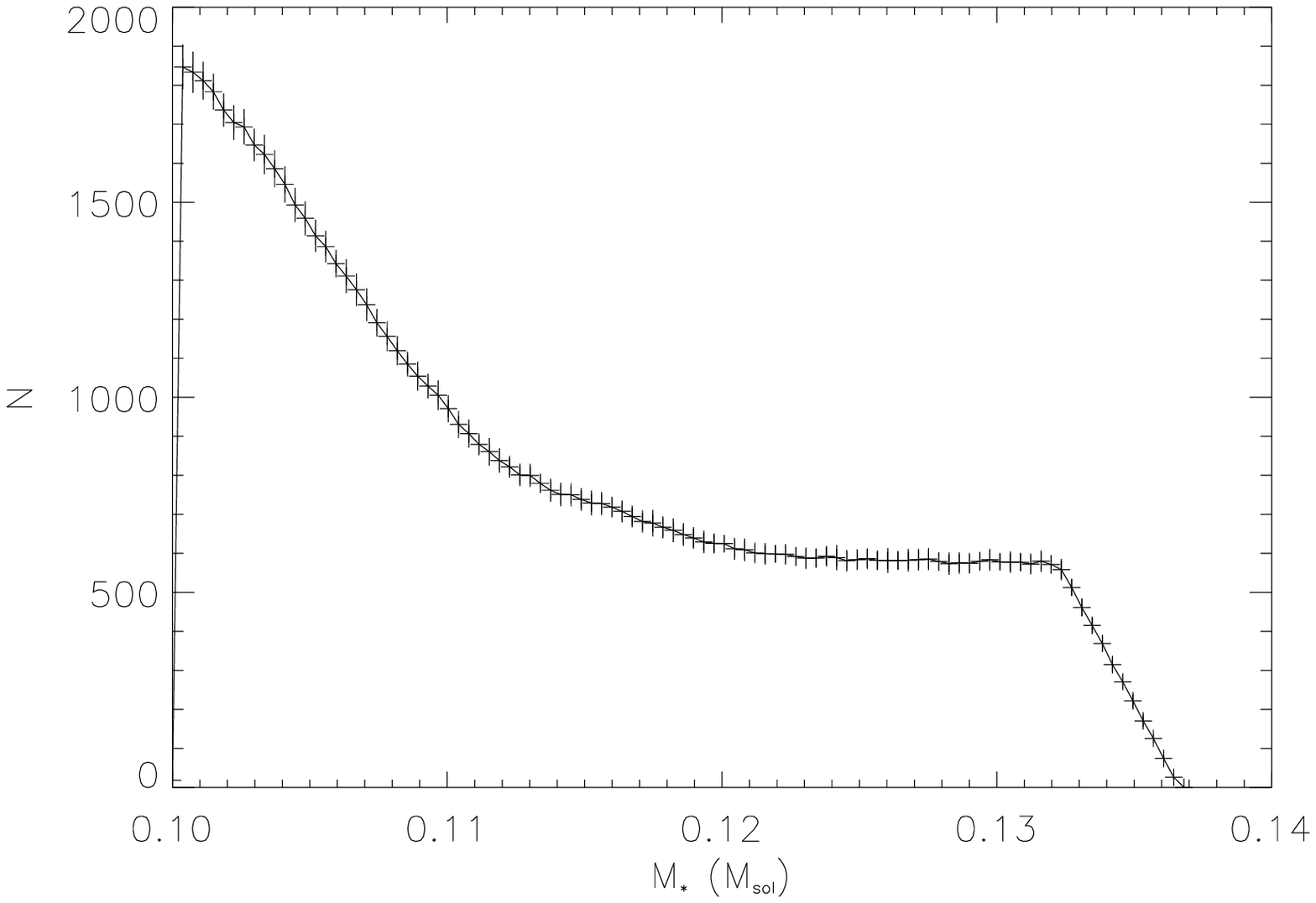} &
\includegraphics[scale = 0.4]{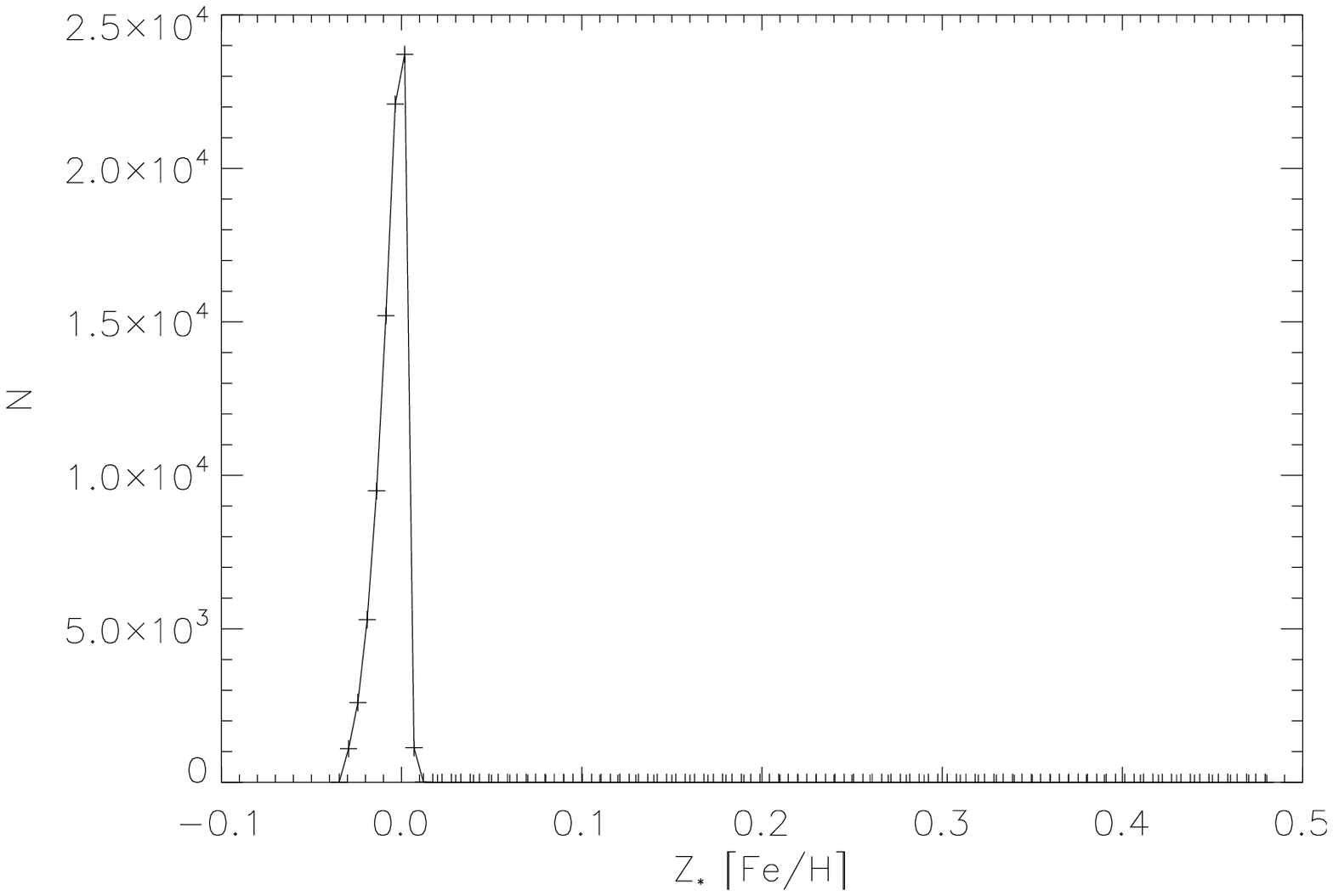} \\
\end{array}$
\caption{\emph{The distribution of host star mass (left) and metallicity (right) under the Tortoise and Hare Hypothesis.} \label{fig:mstar_hare}}
\end{figure}

\begin{figure}
$
\begin{array}{cc}
\includegraphics[scale = 0.4]{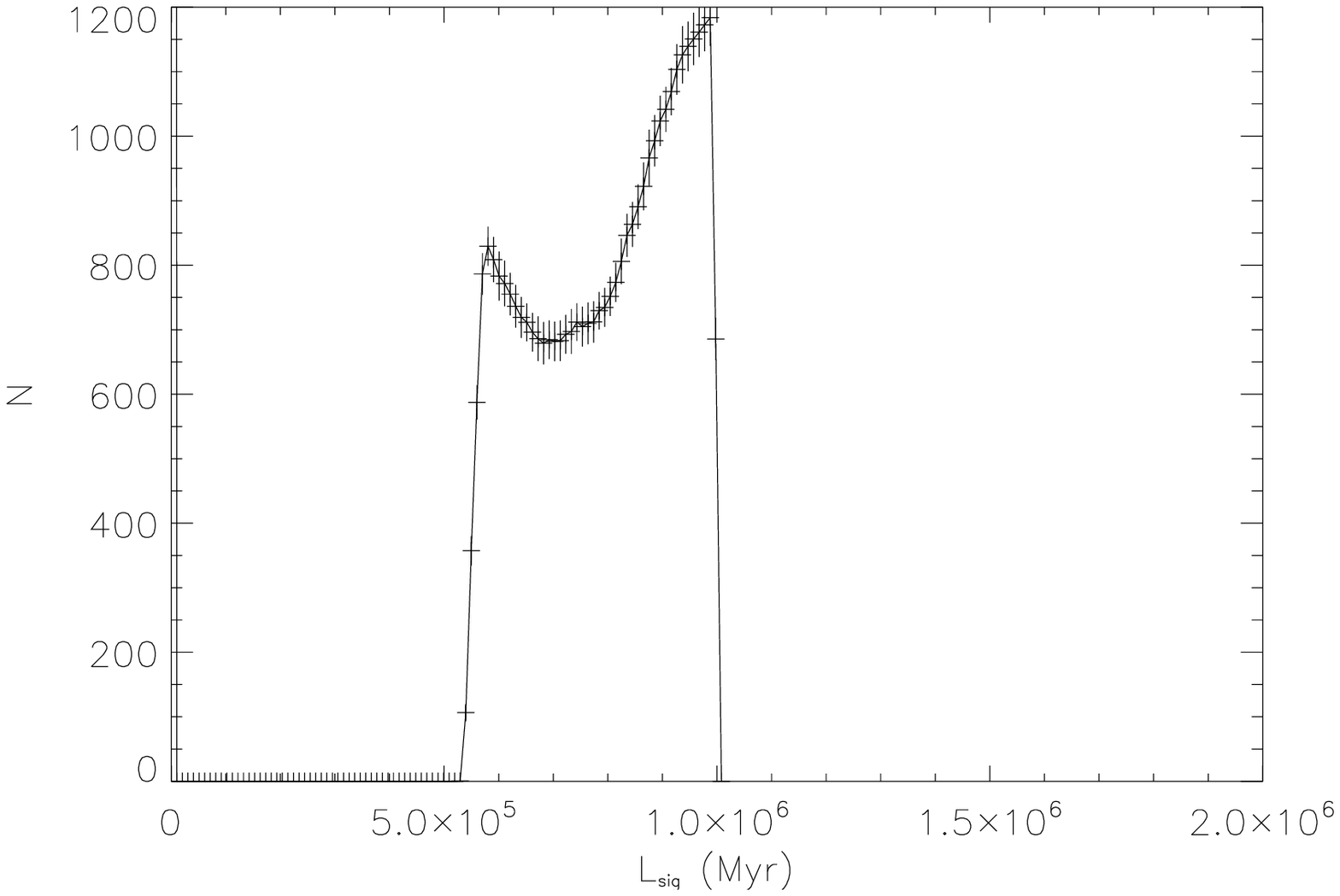} &
\includegraphics[scale = 0.4]{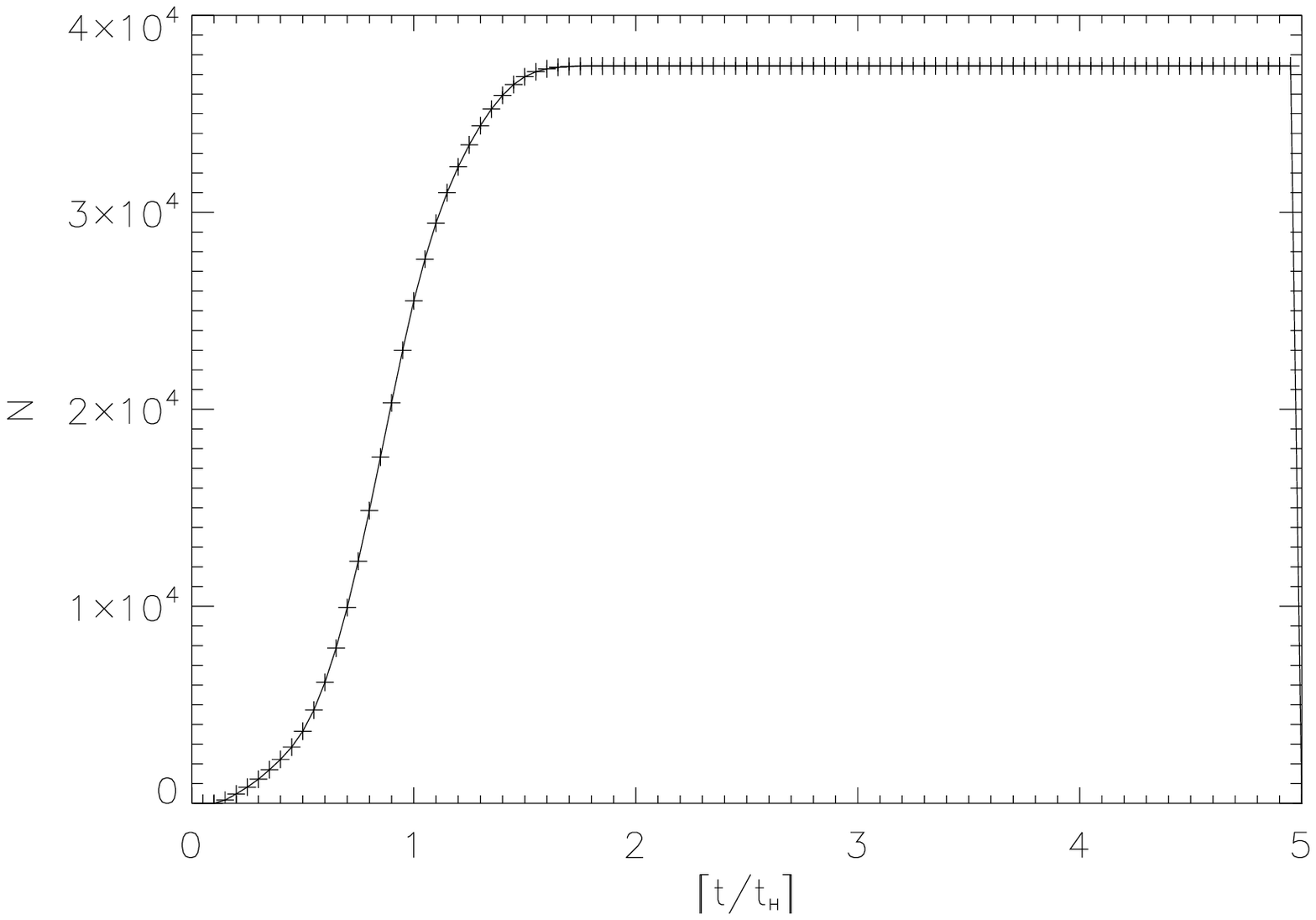} \\
\end{array}$
\caption{\emph{The distribution of signal lifetimes (left) and the signal history of the Galaxy (right) under the Tortoise and Hare Hypothesis.} \label{fig:Lsig_hare}}
\end{figure}

\section{Conclusions}\label{sec:Conclusions}

\noindent The results for all three hypotheses show clear trends throughout.  In all cases, the inhabited planets orbit low mass stars (a symptom of the Hot Jupiter bias present in current data); the signal lifetimes are correspondingly dependent on these masses (as expected, given its functional form).  The signal history of all three hypotheses shows a period of transition between low signal number and high signal number at around \(t \sim t_H\), again symptomatic of the Biological Copernican Principle used to constrain \(\tau_i\), \(N_{stages}\), etc. \\

\noindent This paper has outlined a means by which key SETI variables can be estimated, taking into account the diverse planetary niches that are known to exist in the Milky Way and the stochastic evolutionary nature of life, as well as providing estimates of errors on these variables.  However, two notes of caution must be offered: 

\begin{enumerate}
\item The reader may be suspicious of the high precision of the statistics quoted: it is worth noting that the standard deviations of these results are indeed low, and the data is precise: but, its accuracy is not as certain.  The output data will only be as useful as the input data will allow (the perennial ``garbage in, garbage out'' problem).  Current data on exoplanets, while improving daily, is still insufficient to explore the parameter space in mass and orbital radii, and as such all results here are very much incomplete.  Conversely, as observations improve and catalogues attain higher completeness, the efficacy of the Monte Carlo Realisation method improves also.  Future studies will also consider planetary parameters which are sampled as to match current planet formation theory, rather than current observations. \\
\item The method currently does not produce a realistic age metallicity relation (AMR).  Age, metallicity and galactocentric radius are intrinsically linked in this setup: to obtain realistic data for all three self-consistently requires an improved three-dimensional galaxy model which takes into account its various components (the bulge, the bar, etc), as well as the time evolution of the Galaxy.  Future work will attempt to incorporate a more holistic model which allows all three parameters to be sampled correctly.  In particular, future efforts will be able to take advantage of better numerical models for the star formation history (e.g. Rocha-Pinto et al 2000 \cite{Rocha_Pinto}), and the spatial distribution of stars (e.g. Dehnen and Binney 1998 \cite{Dehnen_Binney}).
\end{enumerate}

\noindent Although this paper applies the ``hard step scenario'' to the biological processes modelled, the method itself is flexible enough to allow other means of evolving life and intelligence. The minutiae of how exactly the biological parameters are calculated do not affect the overall concept: this work has shown that it is possible to simulate a realistic backdrop (in terms of stars and planets) for the evolution of ETI, whether it is modelled by the ``hard step'' scenario or by some other stochastic method.  Incorporating new empirical input from the next generation of terrestrial planet finders, e.g. Kepler (Borucki et al 2008 \cite{Kepler}), as well as other astrobiological research into the model, alongside new theoretical input by adding more realistic physics and biology will strengthen the efficacy of this Monte Carlo technique, providing a new avenue of SETI research, and a means to bring many disparate areas of astronomical research together.

\section{Acknowledgements}

\noindent The author would like to thank the referee for valuable comments and suggestions which greatly improved this paper.   The simulations described in this paper were carried out using high performance computing funded by the Scottish Universities Physics Alliance (SUPA).

\end{document}